\documentclass[12pt]{article}
\usepackage{amsmath,amsthm}
\usepackage{amsfonts}
\usepackage{amssymb}
\usepackage{graphicx,psfrag,epsf}
\usepackage{enumerate}
\usepackage{natbib}
\usepackage{bm}
\usepackage{url}
\usepackage{float}
\usepackage{subfig}
\usepackage{setspace}
\usepackage{booktabs}

\usepackage{enumitem}
\newtheorem{theorem}{Theorem}

\newtheorem{proposition}{Proposition}

\usepackage{threeparttable}
\usepackage{multirow}
\usepackage{array}
\usepackage{pifont}
\usepackage{tabularx}
\usepackage{longtable}

\usepackage{tikz}
\usetikzlibrary{
  positioning,
  arrows.meta,
  calc,
  fit,
  backgrounds
}
\usepackage{hyperref}
\hypersetup{colorlinks=true,
citecolor=blue,   
linkcolor=blue,   
urlcolor=blue     
}
\usetikzlibrary{positioning,arrows.meta,calc}
\usepackage{placeins}

\begin{document}

\def\spacingset#1{\renewcommand{\baselinestretch}%
{#1}\small\normalsize} \spacingset{1}

 \title{\bf \textsc{NAFI}: Bayesian Modeling of Acquisition and
 Curriculum Transfer in Digital Learning}
   \author{Guanyu Hu\\
     Department of Statistics \& Probability, Michigan State University\\
     and \\
     Icy Zhang \\
     Department of Educational Psychology, University of Wisconsin-Madison\\
     and \\
     Zhihua Ma\thanks{
 Correspondence should be addressed to Zhihua Ma, mazh1993@outlook.com }\hspace{.2cm} \\
     Department of Statistics, Shenzhen University}
 \date{}
   \maketitle

%

\bigskip
\begin{abstract}
Digital learning environments combine embedded assessments with engagement
traces, allowing learning to be studied across a curriculum. We examine how
students' current response propensities relate to current engagement and
earlier latent states through curriculum pathways, and how these processes
vary across students.
We propose \textsc{NAFI}, a Bayesian \textbf{N}etwork
\textbf{A}utoregressive \textbf{F}unctional \textbf{I}tem response theory model that decomposes
chapter-level response propensity into engagement-associated acquisition and
curriculum-mediated transfer. Acquisition is modeled using profile- and
topic-specific nondecreasing splines with diminishing increments. Transfer
propagates earlier latent states through a directed network based on
pedagogical similarity, allowing nonadjacent chapters to contribute.
Covariate-dependent latent profiles under a balanced-tree stick-breaking
prior capture learner heterogeneity. We establish shape preservation of
the acquisition functions and a finite path representation of direct and
indirect curriculum propagation. Applied to CourseKata data from an
introductory statistics course, \textsc{NAFI} identifies distinct learner
profiles and yields interpretable summaries of acquisition, transfer
persistence, and curriculum cohesion. The results illustrate how the model
can inform personalized instruction and curriculum design while preserving
an associational interpretation of engagement and earlier learning.
\end{abstract}

\noindent%
\vfill

\newpage
\spacingset{1.45}
\section{Introduction}
\label{sec:intro}

Digital learning environments now record, at scale, both students' responses to
assessment items and the behavioral traces of how they engage with instructional
material, including task time, attempts, page activity, and video interactions
\citep{liu2020bayesian,fries2021connections,shah2024economics}. When formative
questions are embedded directly within instructional pages
\citep{hoch2018interactive,black2009formative}, these records provide repeated
evidence of students' developing understanding during instruction rather than
only at its end, and they can be linked to estimates of how long students
engaged with the content \citep{kovanovic2015time}. Such data make it possible
to study learning as a cumulative process, though they raise several challenges
at the same time. Students' knowledge is latent and develops over extended
periods \citep{abbakumov2019measuring}. The content and assessment items change
from one instructional unit to the next \citep{wang2016secondorder}. Students
differ in how they engage and learn \citep{saqr2023longitudinal}. Knowledge
acquired in one part of a course can also provide the foundation for concepts
encountered much later \citep{fries2021connections}.

This paper is motivated by embedded assessment data from CourseKata, a popular
interactive online textbook for introductory statistics and data science
\citep{stigler2020better}. 
For nine chapters in the textbook, students answered
231 embedded items interleaved with textbook instruction, while their chapter
level engagement was recorded, and five presurvey characteristics were measured
via students' self-reports in a textbook pre-survey before instruction began.
The educational question is how students develop understanding as they engage
with new material and build on knowledge from earlier related chapters, and how
these processes differ across learners. Answering this question requires a
model that separates three sources of variation in performance: learning
associated with engagement in the current chapter, the carryover of earlier
learning through the structure of the curriculum, and systematic differences
among students in both. In this paper, acquisition refers to the component of
chapter-level latent response propensity associated with engagement in the
current chapter. Transfer refers to curriculum-structured dependence of the
current latent state on earlier chapter latent states. Both are model-based
associations.

Item response theory (IRT) gives a principled framework for measuring latent
proficiency while accounting for item difficulty
\citep{Rasch1960,lord1968statistical,baker2001basics,cai2016item}, though it
generally treats ability as fixed over the assessment period. Dynamic and
longitudinal IRT models let ability change through state space, autoregressive,
or related structures
\citep{dunson2006bayesian,hickendorff2009solution,wang2013dynamic,jeon2016autoregressive,liu2020bayesian,liu2025general,sun2025bayesian}.
In most of these formulations, the current state depends on the most recent
state or follows a prespecified transition. Closer to the present work,
\citet{tamano2023dynamical} extend multidimensional IRT to the longitudinal
setting, letting several latent skills evolve through a linear dynamical system
while combining non-compensatorily within each item, with a variational
approximation and Monte Carlo EM for inference. Our model differs in what the
latent state represents and in how dependence is structured. The latent state
is a single chapter-level response propensity measured on a common Rasch
scale, so proficiency is comparable across chapters with no anchor items, and
dependence propagates through a directed curriculum network built from
pedagogical similarity rather than a first-order Markov transition, which
allows nonadjacent chapters to contribute and yields a finite path
representation of direct and indirect transfer. The model also separates
acquisition tied to engagement from transfer of earlier proficiency, a
decomposition that dynamical MIRT formulations do not consider. Models from
learning analytics take a complementary view. Knowledge tracing tracks mastery
of predefined skills \citep{corbett1995knowledge,piech2015deep}. Additive and
performance factors models relate success to counts of prior practice
opportunities on each skill \citep{cen2006learning,pavlik2009performance}.
Hidden Markov cognitive diagnosis models describe transitions among discrete
attribute profiles \citep{chen2018hidden,wang2018tracking}. These approaches
model learning from practice within skills. They do not, in general, separate
learning associated with engagement in the current unit from the carryover of
earlier proficiency through a curriculum network, while also accounting for
differences in item difficulty across units that share no anchor items.
A second challenge is that students with similar observed performance may
arrive there in different ways. Some may struggle because earlier knowledge
does not carry over, while others may build on earlier knowledge effectively
but acquire new material more slowly. Differential item functioning,
multilevel IRT, mixture IRT, and latent transition models provide important
tools for modeling population heterogeneity
\citep{taple2022impact,fox2005multilevel,cho2010latent,nylund2023ten,pan2024bayesian,hu2024nonparametric}.
These methods, however, generally neither separate these learning mechanisms
nor link the resulting latent groups directly to observed learner
characteristics.

We introduce \textsc{NAFI}, a Bayesian \textbf{N}etwork \textbf{A}utoregressive
\textbf{F}unctional \textbf{I}tem response theory model for chapter-level
embedded assessment data. It pairs a Rasch type measurement model with a latent
process for each student's chapter specific response propensity, built from
three components. First, within each chapter, the effect of current engagement
is modeled by profile and topic specific B-spline curves that are nondecreasing
with decreasing coefficient increments. This reflects the diminishing returns
to practice described in cognitive psychology \citep{anderson1982acquisition}.
Shape restricted regression imposes such scientifically motivated constraints
while staying nonparametric \citep{meyer2008inference,pya2015shape}, and it
prevents arbitrary oscillation where engagement is sparsely observed. We treat
these restrictions as modeling assumptions. Second, earlier response propensity feeds
into the current chapter through a directed lower triangular curriculum network
built from pedagogical similarity between chapters, so nonadjacent chapters can
contribute directly. Here, transfer means how far proficiency developed in an
earlier unit carries into performance in a later one, beyond new acquisition in
the later unit. This is narrower than transfer in the psychology of learning,
which concerns applying knowledge in new contexts \citep{barnett2002when}, and
we interpret it accordingly. Third, students belong to latent profiles, each
with its own acquisition curves and transfer persistence, and profile
membership depends on presurvey characteristics. Bayesian CART
\citep{chipman1998bayesian,rockova2020posterior} could partition students
through recursive covariate splits and give readable decision rules, but its
tree search moves complicate implementation and can mix poorly. We instead use
the tree stick-breaking construction of \citet{horiguchi2022tree}, a covariate
dependent extension of logistic stick-breaking \citep{ren2011logistic} that
keeps P{\'o}lya-Gamma augmentation \citep{polson2013bayesian}. Its balanced
topology gives every profile a path of length $\log_2 K$, avoiding the
persistent positive correlation among profile probabilities and the chain
reaction label switching of the usual lopsided construction. Profiles are
identified only by ordering their baseline locations. Transfer persistence is
unconstrained, so the data, not the prior, decide whether stronger baseline
performance goes with stronger transfer.

For educational assessment, \textsc{NAFI} shifts latent proficiency measurement
from a static account of what students know toward a decomposition of how
performance develops across interrelated chapters. Methodologically, the model
unites shape constrained functional acquisition, network structured dependence
among latent states, and covariate dependent mixture weights in one Bayesian
hierarchy. We show that, because the network is acyclic, the transfer process
has a finite path representation, which gives interpretable summaries of direct
and indirect propagation between chapters and of profile specific curriculum
cohesion. Applied to the CourseKata data, the model separates students whose
difficulties come mainly from limited acquisition from those whose earlier
learning carries over weakly, and it recovers distinct learner profiles linked
to presurvey characteristics. Because the curriculum network is fixed in
advance, the profile-specific transfer parameters also show how strongly the
designed curriculum's pathways propagate earlier proficiency for different
groups of learners. From a statistical perspective, \textsc{NAFI} provides a general
framework for jointly modeling heterogeneous functional trajectories,
structured dependence among latent processes, and latent population
heterogeneity, with interpretable measures of how information
propagates through a network. From an educational perspective,
the framework moves beyond conventional proficiency scores by
distinguishing knowledge acquisition from knowledge transfer,
providing a more detailed understanding of how students learn
across connected curricular topics. For the design and evaluation
of teaching experiments, these distinctions offer a principled
basis for identifying students who may benefit from different
instructional strategies and for developing more targeted
interventions, whose effectiveness can subsequently be
evaluated through randomized or appropriately designed
quasi-experimental studies.

\section{Motivating Data and Educational Questions}
\label{sec:motivation}

Learning is cumulative: new knowledge builds on and connects with what
students already understand. In the \textit{knowledge-in-pieces}
perspective, learners draw on diverse, context-sensitive knowledge
resources; developing understanding involves reorganizing these resources
and refining when and how they are activated
\citep{disessa1993epistemology,disessa2014causal}. The \textit{practicing connections} framework
emphasizes that coherent understanding develops as learners connect concepts,
representations, and contexts over time \citep{fries2021connections}. Lastly, the
\textit{knowledge integration} perspective similarly characterizes
learning as a process of distinguishing, evaluating, and connecting
existing and newly encountered ideas \citep{linn2011science}. 
There is a strong theoretical foundation for the proposition that earlier experiences
can support learners' capacity to interpret and learn from subsequent
instruction, beyond their ability to apply an already learned procedure
\citep{bransford1999transfer}. In mathematics, experimental evidence
further suggests that conceptual understanding and procedural skill can
develop iteratively, with advances in one supporting advances in the
other \citep{rittlejohnson2001iterative}. 
Together, these accounts motivate
examining learning across conceptually related parts of a curriculum.
Understanding in a later chapter may reflect both engagement with current
material and resources developed in earlier chapters, with the relevance
of those resources depending on the conceptual connections between them.
Curriculum order alone, however, does not establish that students have
formed or used those connections.

Importantly, the development and use of knowledge may also vary across learners. For example, situated expectancy-value theory links students' expectations of success
and subjective task values to achievement-related choices, persistence,
and performance \citep{eccles2020situated}. Control-value theory explains
how perceived control and value shape achievement emotions and how these
emotions relate reciprocally to engagement and achievement
\citep{pekrun2006control}. Mathematics anxiety may additionally interfere
with working-memory resources available during mathematical performance
\citep{ashcraft2007working}. Self-regulated learning theory emphasizes
learners' goal setting, monitoring, strategy use, and adaptation of their
learning activities \citep{zimmerman2002selfregulated}. These accounts
identify complementary reasons why students with similar initial
knowledge may differ in how they engage with new material and use earlier
understanding thus motivate examining how variation in these learner characteristics might influence the learning process.

\begin{figure}[!htbp]
\centering
\captionsetup[subfloat]{font=scriptsize}
\subfloat[Current acquisition]{
\includegraphics[scale=0.35]{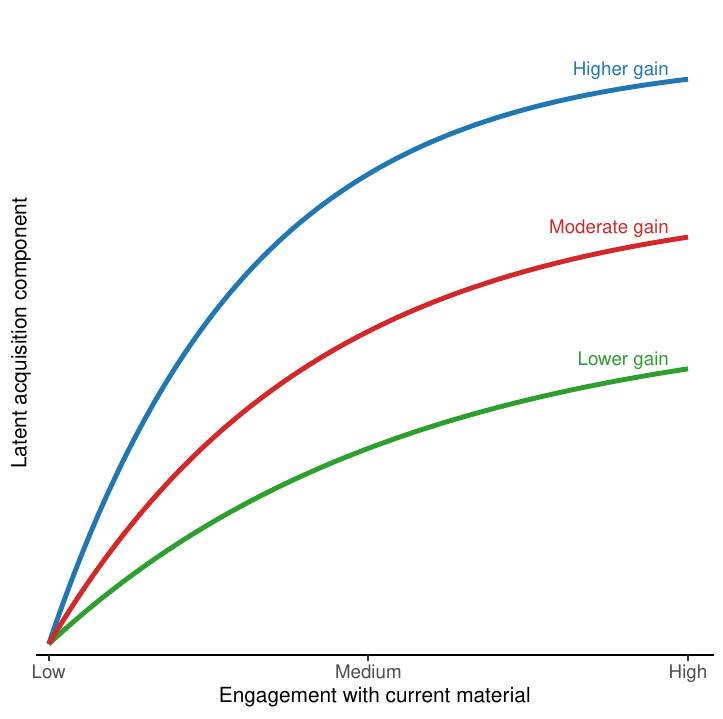}}
\subfloat[Curriculum-mediated transfer]{
\includegraphics[scale=0.35]{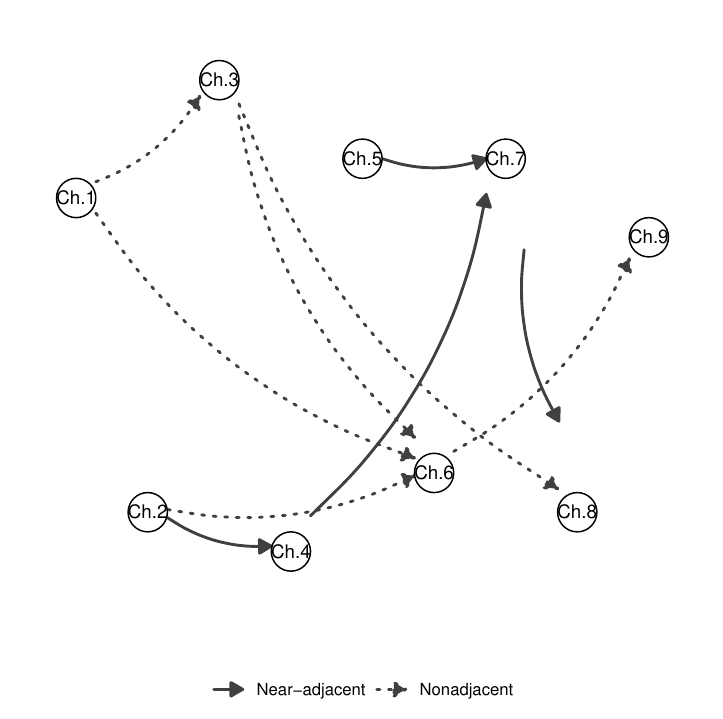}}
\subfloat[Learner heterogeneity]{
\includegraphics[scale=0.35]{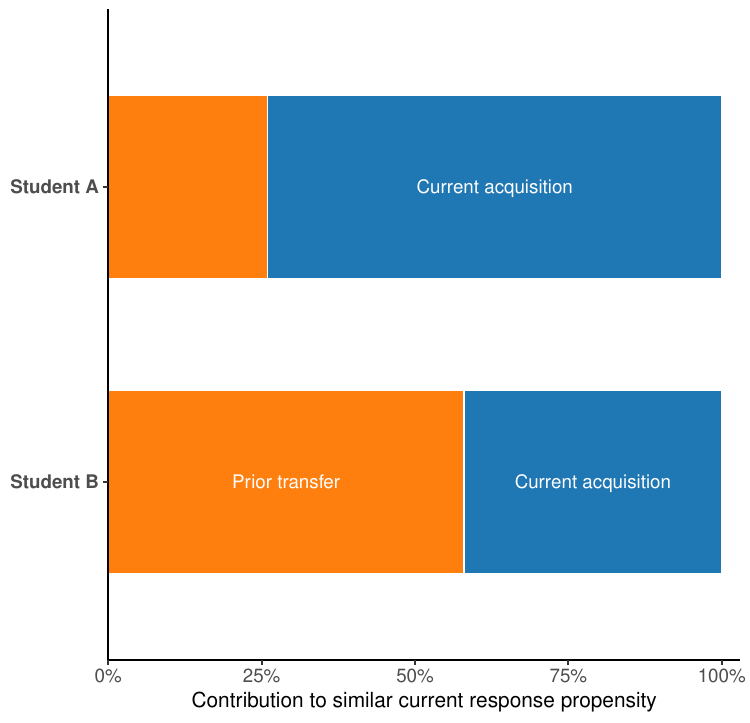}}
\caption{
\textbf{Learning through acquisition and curriculum-mediated transfer.}
(a) Engagement with current material is associated with new acquisition, with
the possibility of diminishing marginal gains.
(b) Previously acquired knowledge can contribute to later learning through
pedagogically related parts of the curriculum, including nonadjacent chapters.
(c) Students with similar observed performance may differ in the relative
contributions of current acquisition and prior-knowledge transfer.
}
\label{fig:motivation}
\end{figure}

Digital learning environments allow these processes to be examined
longitudinally. Embedded assessments provide repeated evidence of developing
understanding, while interaction logs document learning activities
\citep{hoch2018interactive} and engagement with instructional content
\citep{kovanovic2015time}. Linking these records across chapters and to
presurvey characteristics makes it possible to study how knowledge
development varies across students.
Together, these considerations motivate our central educational question:
\begin{quote}
\emph{How do students develop understanding as they engage with new
material and build on knowledge acquired in earlier pages and
chapters, and how do these processes vary with learner characteristics?}
\end{quote}
Addressing this question requires a model of engagement-associated
acquisition, contributions from earlier related chapters, and variation in
these processes across learners. Figure~\ref{fig:motivation} illustrates
these three features: current acquisition, knowledge building across
chapters, and learner heterogeneity.

CourseKata interleaves instructional text, conceptual questions,
and executable R exercises, allowing students to read, practice, and receive
feedback in one environment \citep{stigler2020better,sutter2026perceptions}.
Its data infrastructure records item responses and content interactions and
links them to precourse surveys of students' prior experience and
psychological characteristics \citep{stigler2020better,zheng2025linking}.
We use data collected during Winter 2023 from undergraduates enrolled in an
introductory statistics and data science course at a North American
university \citep{CourseKata2023}. Of the 221 students who began the course,
198 completed at least one item in every chapter and were retained for the
primary analysis.

The curriculum contained $T=9$ chapters, with students answering questions
continuously rather than completing a single examination after each unit.
The chapter-specific item counts were
\[
(J_1,\ldots,J_9)
=
(13,32,31,32,32,30,12,20,29),
\]
yielding 231 binary items for a complete student record before accounting
for missing responses. This repeated structure provides finer information
about performance during instruction than a small number of summative
assessments. For each student and chapter, we combine activity counts,
activity duration, and an engagement ratio into a chapter-level engagement
measure $x_{it}$.

Five presurvey characteristics are available: biological sex, high
mathematics anxiety, belief that statistics is learned primarily through
memorization, strong prior mathematics performance, and previous coding
experience. Because these characteristics were measured before instruction,
they allow us to examine whether student backgrounds are associated with
different acquisition and knowledge-transfer patterns.

A course design expert specified our knowledge-transfer network to encode
the course's intended pedagogical and prerequisite connections. The network
identifies earlier chapters that provide conceptual foundations for later
material, including nonadjacent chapters, rather than assuming that knowledge
passes only between consecutive units. We represent these connections using
the nonnegative structure matrix $S_{tc}$, where a positive entry indicates
that earlier chapter $c$ is expected to support later chapter $t$, and
$S_{tc}=0$ for $c\geq t$. The network is fixed before model fitting and is
not learned from student responses. Its purpose is to incorporate curriculum
design into the distinction between current-chapter acquisition and the
contribution of earlier learning through direct and indirect pathways.

\begin{figure}[!htbp]
\centering
\subfloat[Performance]{
\includegraphics[scale=0.4]{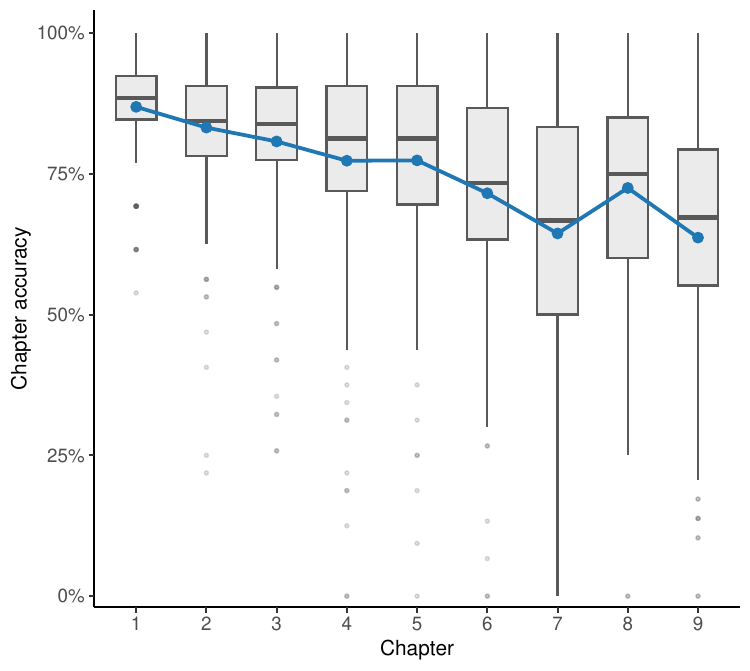}}
\subfloat[Engagement]{
\includegraphics[scale=0.4]{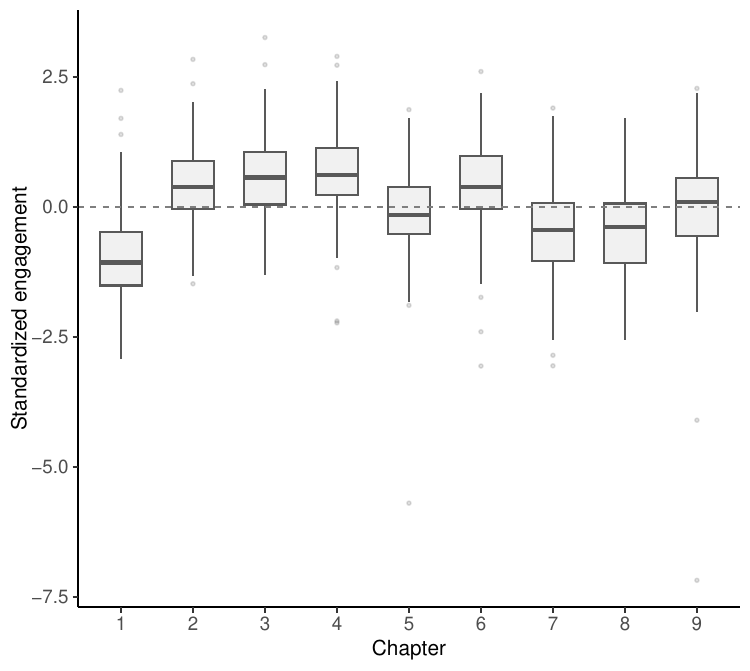}}

\subfloat[Curriculum structure]{
\includegraphics[scale=0.4]{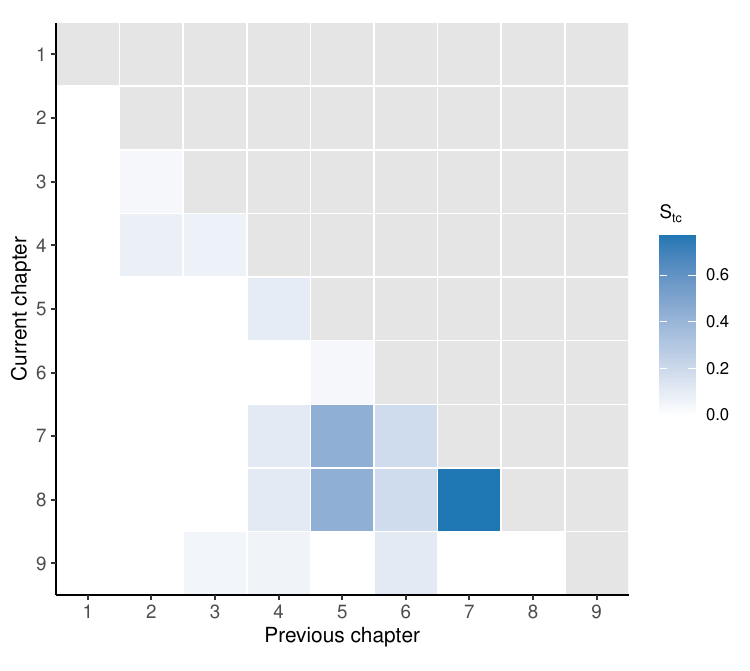}}
\subfloat[Student heterogeneity]{
\includegraphics[scale=0.4]{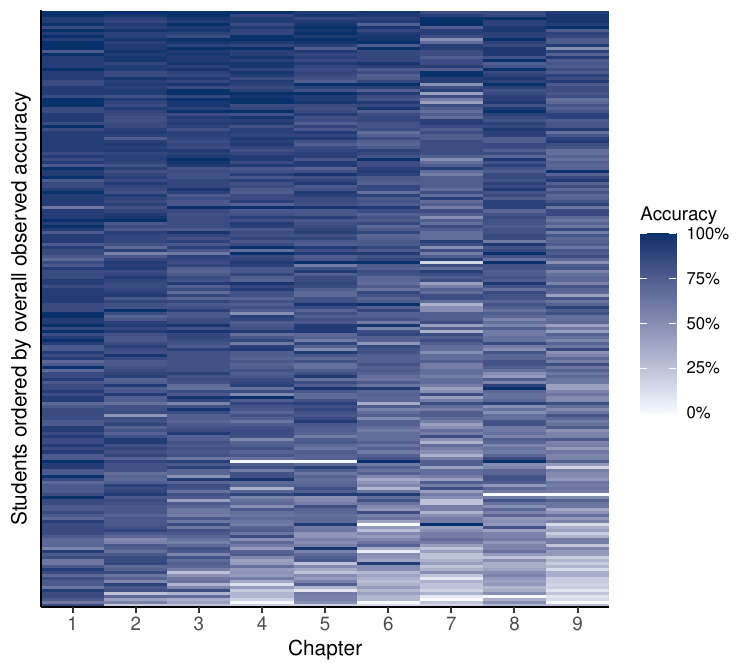}}
\caption{
\textbf{Features of the motivating CourseKata data.}
(a) Assessment performance vary across instructional chapters.
(b) Engagement varies across instructional chapters. 
(c) The curriculum similarity structure includes relationships between
nonadjacent chapters, suggesting that temporal proximity alone does not capture
the relevant learning history.
(d) Students exhibit heterogeneous longitudinal performance patterns across the
curriculum.
}
\label{fig:coursekata}
\end{figure}

Figure~\ref{fig:coursekata} summarizes the empirical motivation. Accuracy
and engagement vary across chapters, the expert-designed network contains
nonlocal pedagogical relationships, and the student-by-chapter accuracy
heatmap reveals heterogeneous longitudinal trajectories. Together, these
features motivate a model that accounts for item difficulty, curriculum
structure, and student-level variation.

%
%
%
%
%
%
%
%
%
%

The CourseKata data provide repeated item-level measurements while students
are actively learning, together with information on engagement, curriculum
structure, and presurvey characteristics. This setting raises questions that
are difficult to address using either static IRT models or conventional
first-order longitudinal models. Static IRT provides a principled framework
for accounting for differences in item difficulty but does not describe how
latent knowledge develops during instruction. Dynamic IRT and knowledge-tracing
models allow knowledge to change over time
\citep{wang2013dynamic,wang2020longitudinal,corbett1995knowledge}, but they
typically do not distinguish learning associated with current engagement from
the contribution of earlier, pedagogically related material.

These features motivate three scientific questions that guide our methodological
development and subsequent analysis:
\vspace{-6pt}
\begin{enumerate}[label=({Q}\arabic*)]

    \item \textit{How is engagement with current material associated with new
    knowledge acquisition?}
    Additional engagement need not correspond to a constant amount of learning.
    We therefore seek to characterize how acquisition changes over the observed
    engagement range, including whether marginal gains become smaller as
    engagement accumulates, and whether these patterns differ across
    pedagogical topics.
    \vspace{-6pt}

    \item \textit{How does knowledge developed earlier contribute to learning
    later in the curriculum?}
    Concepts introduced in one chapter may become useful again several chapters
    later, so temporal proximity alone may not capture the relevant learning
    history. We therefore ask how knowledge propagates through pedagogically
    related chapters, including both direct connections and indirect pathways
    through intermediate material.
    \vspace{-6pt}

    \item \textit{Do students differ in how they acquire new knowledge and
    build on earlier learning?}
    Students with similar observed performance may arrive there through
    different learning histories. We therefore ask whether students exhibit
    distinct combinations of engagement-related acquisition and curriculum
    transfer, and whether presurvey characteristics are associated with these
    different patterns of knowledge development.

\end{enumerate}
\vspace{-6pt}

Together, these questions move the analysis beyond observed chapter
performance by separating current acquisition from curriculum-mediated
transfer while accounting for item difficulty and learner heterogeneity.
A central challenge is that successive chapters use different items whose
difficulty may not be comparable. Chapter accuracy may therefore reflect
changes in knowledge, assessment content, or both. Jointly modeling item
difficulty and latent knowledge development clarifies students'
trajectories and can support evaluations of interventions, such as videos
embedded across chapters, whose benefits may emerge gradually through
engagement and earlier learning.
These trajectories distinguish limited gains despite sustained engagement
from more rapid acquisition. This distinction is consistent with the ICAP
framework's emphasis on how cognitive engagement supports knowledge
construction \citep{chi2014icap}, although engagement time alone does not
identify those processes. Gains may also remain localized or contribute
broadly to later understanding, reflecting the practicing connections
framework's emphasis on connections among concepts
\citep{fries2021connections}. An intervention extension could examine
whether these patterns emerge in particular chapters, persist, or vary
across learner profiles.
This focus on individual trajectories aligns with precision education
\citep{hart2016precision} and personalized approaches based on repeated
measurement \citep{tetzlaff2021personalized}. Linking presurvey
characteristics to learning profiles describes how students with different
backgrounds may develop knowledge differently. These interpretations remain
conditional on the measurement model: item-difficulty adjustment alone does
not establish comparability across unlinked item sets. Moreover, our
observational analysis does not identify causal effects of engagement or
earlier learning; intervention effects require an appropriate experimental
or causal design. The profiles therefore summarize learning patterns and
generate hypotheses rather than define fixed student types or directly
measure cognitive mechanisms.

\section{The Proposed Model}
\label{sec:model}

\begin{figure}[!htbp]
\centering

\definecolor{modelred}{RGB}{190,35,45}
\definecolor{modelblue}{RGB}{55,115,185}
\definecolor{paramyellow}{RGB}{252,239,190}
\definecolor{hyperblue}{RGB}{155,180,220}
\definecolor{inputgray}{RGB}{235,235,235}

\resizebox{\textwidth}{!}{
\begin{tikzpicture}[
    >=Latex,
    model/.style={
        draw=modelred,
        line width=1.0pt,
        fill=white,
        align=center,
        inner sep=5pt,
        font=\small
    },
    param/.style={
        draw=black!65,
        fill=paramyellow,
        align=center,
        minimum width=17mm,
        minimum height=8mm,
        inner sep=3pt,
        font=\small
    },
    hyper/.style={
        draw=black!65,
        fill=hyperblue,
        align=center,
        minimum width=17mm,
        minimum height=8mm,
        inner sep=3pt,
        font=\small
    },
    input/.style={
        draw=black!60,
        fill=inputgray,
        align=center,
        minimum width=20mm,
        minimum height=8mm,
        inner sep=3pt,
        font=\small
    },
    blackarrow/.style={
        ->,
        line width=0.8pt,
        black,
        shorten >=1pt,
        shorten <=1pt
    },
    bluearrow/.style={
        ->,
        line width=1.0pt,
        modelblue,
        shorten >=1pt,
        shorten <=1pt
    },
    route/.style={
        rounded corners=4pt
    }
]


\node[model, text width=92mm] (irt) at (0,5.0) {
\[
Y_{ijt}\mid\theta_{it},b_{jt}
\sim
\operatorname{Bernoulli}(p_{ijt}),
\qquad
\operatorname{logit}(p_{ijt})
=
\theta_{it}-b_{jt}
\]
};

\node[model, text width=86mm] (acq) at (-4.8,1.15) {
\[
f_{km}(x)
=
\sum_{\ell=1}^{L}
B_{\ell}(x)\beta_{km\ell}
\]
\vspace{-0.18cm}
\[
\beta_{km1}
=
\beta^{(0)}_{km},
\qquad
\beta^{(0)}_{km}
\sim
N(a_k+c_m,\sigma_{\beta0}^2)
\]
\vspace{-0.18cm}
\[
\beta_{km,\ell+1}
=
\beta_{km\ell}
+
\Delta_{km\ell},
\qquad
\sum_{\ell=1}^{L-1}
\Delta_{km\ell}
=
G_{km}
\]
};

\node[model, text width=76mm] (dynamic) at (4.8,1.15) {
\[
\theta_{it}
\sim
N(\mu_{it},\sigma_{\epsilon}^2)
\]
\vspace{-0.18cm}
\[
\mu_{i1}
=
f_{z_i,m(1)}(x_{i1})
+
\delta_1
\]
\vspace{-0.10cm}
\[
\mu_{it}
=
f_{z_i,m(t)}(x_{it})
+
\lambda_{z_i}
\sum_{c<t}W_{tc}\theta_{ic}
+
\delta_t,
\quad t\ge 2
\]
};

\node[model, text width=102mm] (profile) at (0.2,-3.55) {
\[
z_i
\sim
\operatorname{Categorical}
(\phi_{i1},\ldots,\phi_{iK})
\]
\vspace{-0.18cm}
\[
\phi_{ik}
=
\prod_{\ell=1}^{d}
V_{i,r_\ell(k)}^{1-q_\ell(k)}
\{1-V_{i,r_\ell(k)}\}^{q_\ell(k)},
\qquad
V_{ir}
=
\operatorname{logit}^{-1}
(\mathbf w_i^\top\boldsymbol\gamma_r)
\]
};


\draw[bluearrow]
(profile.north) -- (acq.south east);

\draw[bluearrow]
(profile.north) -- (dynamic.south west);

\draw[bluearrow]
(acq.east) -- (dynamic.west);

\draw[bluearrow]
(dynamic.north) -- (irt.south east);


\begin{scope}[on background layer]
\node[
    draw=modelred,
    dashed,
    line width=1.1pt,
    inner xsep=7mm,
    inner ysep=2.5mm,
    fit=(irt)(acq)(dynamic)(profile)
] (modelbox) {};
\end{scope}


\node[hyper] (sigmab) at (-4.8,7.0) {$\sigma_{b,t}$};
\node[above=1mm of sigmab,font=\scriptsize]
{$\sim\operatorname{HN}(0,2^2)$};

\node[param] (braw) at (-2.4,7.0) {$b^{\rm raw}_{jt}$};
\node[above=1mm of braw,font=\scriptsize]
{$\sim N(0,\sigma_{b,t}^2)$};

\node[param] (b) at (0,7.0) {$b_{jt}$};
\node[above=1mm of b,font=\scriptsize]
{chapter-centered};

\node[hyper] (sigmachapter) at (6.2,7.0) {$\sigma_{\delta}$};
\node[above=1mm of sigmachapter,font=\scriptsize]
{$\sim\operatorname{HN}(0,0.35^2)$};

\node[param] (alpha) at (8.8,7.0) {$\delta_t$};
\node[above=1mm of alpha,font=\scriptsize]
{$\sum_{t=1}^{T}\delta_t=0$};

\draw[blackarrow] (sigmab) -- (braw);
\draw[blackarrow] (braw) -- (b);

\draw[blackarrow]
(b.south) -- (irt.north);

\draw[blackarrow] (sigmachapter) -- (alpha);

\draw[blackarrow]
(alpha.south) -- (dynamic.north east);


\node[input] (xinput) at (-9.3,7.0)
{$x_{it}$};
\node[above=1mm of xinput,font=\scriptsize]
{standardized engagement};

\node[param] (a) at (-11.25,5.00)
{$a_k$};
\node[above=1mm of a,font=\scriptsize]
{$a_1>\cdots>a_K$};

\node[param] (abase) at (-14.0,5.75)
{$a_1$};
\node[above=1mm of abase,font=\scriptsize]
{$\sim N(a_1^\star,0.35^2)$};

\node[param] (agap) at (-14.0,4.60)
{$g^a_k$};
\node[below=1mm of agap,font=\scriptsize]
{$a_{k+1}=a_k-g^a_k$};

\node[input] (c) at (-11.25,2.70)
{$c_m$};
\node[above=1mm of c,font=\scriptsize]
{fixed topic anchor};

\node[hyper] (sigma0) at (-11.25,1.15)
{$\sigma_{\beta0}$};
\node[above=1mm of sigma0,font=\scriptsize]
{$\sim\operatorname{HN}(0,0.5^2)$};

\node[param] (G) at (-11.2,-2.5)
{$G_{km}$};
\node[below=0.2mm of G,font=\scriptsize]
{$\sim N(G^\star_{km},\sigma_G^2)_{+}$};

\node[hyper] (sigmag) at (-11.2,-1.1)
{$\sigma_G$};
\node[above=1mm of sigmag,font=\scriptsize]
{$\sim\operatorname{HN}(0,1.00^2)$};

\node[param] (rho) at (-13.2,-2.5)
{$\rho_{km\ell}$};
\node[above=1mm of rho,font=\scriptsize]
{$\sim\operatorname{Beta}(4,2)$};

\node[model, text width=74mm] (delta) at (-9.3,-6.2) {
\[
u_{km1}=1,\qquad
u_{km,\ell+1}=u_{km\ell}\rho_{km\ell}
\]
\vspace{-0.06cm}
\[
\Delta_{km\ell}
=
G_{km}
\frac{u_{km\ell}}
{\sum_{r=1}^{L-1}u_{kmr}},
\quad
\ell=1,\ldots,L-1
\]
};

\draw[blackarrow, route]
(xinput.south) -- (acq.north west);

\draw[blackarrow] (abase.east) --  (a.north west);
\draw[blackarrow] (agap.east) --  (a.west);

\draw[blackarrow] (a.east) --  (acq.north west);
\draw[blackarrow] (c.east) -- (acq.west);
\draw[blackarrow] (sigma0.east) --  (acq.west);

\draw[blackarrow] (sigmag) -- (G);
\draw[blackarrow] (G.east) --  (delta.north);
\draw[blackarrow] (rho.south) --(delta.north west);
\draw[blackarrow] (delta.north) --  (acq.south west);


\node[input] (W) at (11.25,3.35)
{$\mathbf W=(W_{tc})$};
\node[above=1mm of W,font=\scriptsize]
{fixed curriculum network};

\node[param] (lambda) at (11.25,1.35)
{$\lambda_k$};
\node[above=1mm of lambda,font=\scriptsize]
{$\eta_k=\operatorname{logit}(\lambda_k)$};
\node[below=1mm of lambda,font=\scriptsize]
{$\eta_k\sim N(\eta_k^\star,0.75^2)$};

\node[hyper] (sigeps) at (11.25,-0.85)
{$\sigma_{\epsilon}$};
\node[below=1mm of sigeps,font=\scriptsize]
{$\sim\operatorname{HN}(0,0.50^2)$};

\draw[blackarrow] (W) -- (dynamic.east);
\draw[blackarrow] (lambda) -- (dynamic.east);
\draw[blackarrow] (sigeps) -- (dynamic.east);


\node[input] (w) at (-1.45,-6.85)
{$\mathbf w_i$};
\node[below=1mm of w,font=\scriptsize]
{presurvey characteristics};

\node[param] (gamma) at (1.8,-6.85)
{$\boldsymbol\gamma_r$};
\node[below=1mm of gamma,font=\scriptsize]
{$\sim N(0,0.5^2)$};

\draw[blackarrow] (w) -- (profile.south);
\draw[blackarrow] (gamma) -- (profile.south);


\node[param, minimum width=11mm, minimum height=5mm]
(leg1) at (10.3,-6.05) {};

\node[right=1mm of leg1,font=\scriptsize]
{model parameter};

\node[hyper, minimum width=11mm, minimum height=5mm]
(leg2) at (10.3,-6.90) {};

\node[right=1mm of leg2,font=\scriptsize]
{hyperparameter};

\node[input, minimum width=11mm, minimum height=5mm]
(leg3) at (10.3,-7.75) {};

\node[right=1mm of leg3,font=\scriptsize]
{observed or fixed input};

\end{tikzpicture}
}

\caption{
\textbf{Hierarchical representation of the proposed Bayesian
network-functional dynamic item response model.}
The central dashed region contains the four main components of the model:
Rasch-type item-response measurement, profile- and topic-specific
engagement-acquisition curves, curriculum-mediated dynamic transfer, and
covariate-dependent learner profiles. 
Black arrows connect
parameters and observed or fixed inputs to their corresponding model
components, while blue arrows represent dependencies among latent components.
}
\label{fig:current_model}
\end{figure}
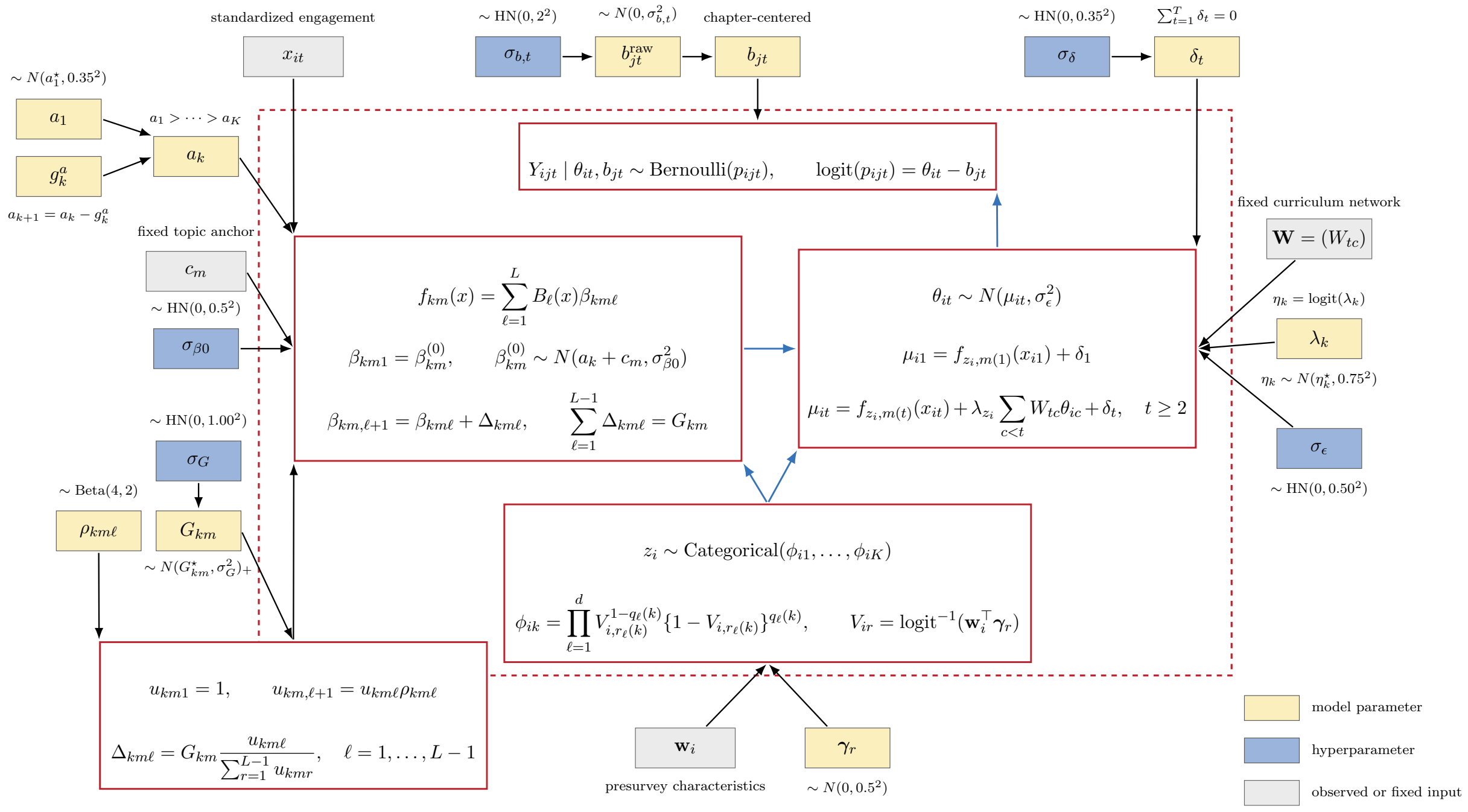

\textsc{NAFI} combines Rasch-type measurement, profile- and topic-specific acquisition curves, curriculum-network transfer, and covariate-dependent learner profiles. Figure~\ref{fig:current_model} summarizes the hierarchy. Let $i=1,\ldots,N$ index students, $t=1,\ldots,T$ chapters, and $j=1,\ldots,J_t$ items within chapter $t$; $m(t)\in\{1,\ldots,M\}$ denotes the pedagogical topic of chapter $t$.

\subsection{Measurement and Covariate-Dependent Learner Profiles}
\label{subsec:measurement_profiles}

For binary response $Y_{ijt}$, the Rasch-type measurement model is
\begin{equation}
Y_{ijt}\mid\theta_{it},b_{jt}
\sim \operatorname{Bernoulli}(p_{ijt}),
\qquad
\operatorname{logit}(p_{ijt})=\theta_{it}-b_{jt},
\label{eq:irt}
\end{equation}
where $\theta_{it}$ is the chapter-specific latent response propensity and $b_{jt}$ is item difficulty. We avoid estimating separate discrimination parameters from the relatively small, non-overlapping chapter item sets. Difficulties are centered by
\[
b_{jt}=b^{\mathrm{raw}}_{jt}-\frac{1}{J_t}\sum_{h=1}^{J_t}b^{\mathrm{raw}}_{ht},
\qquad \sum_{j=1}^{J_t}b_{jt}=0.
\]
Because chapters contain different, unlinked items, cross-chapter comparisons rely on the acquisition, chapter-effect, and transfer constraints as well as centering. Thus, $\theta_{it}$ is a model-based response propensity, not a directly observed common-scale ability.

\paragraph{Learner profiles.}
Let $z_i\in\{1,\ldots,K\}$ denote a latent profile, with $K$ is the number of distinct profiles. A balanced tree of depth $d=\log_2K$ links profile membership to $\mathbf w_i=(1,w_{i2},\ldots,w_{iP})^\top$. At node $r$, the left-branch probability is $V_{ir}=\operatorname{logit}^{-1}(\mathbf w_i^\top\boldsymbol\gamma_r)$. With path nodes $r_\ell(k)$ and branch indicators $q_\ell(k)$,
\begin{equation}
\phi_{ik}
=\prod_{\ell=1}^{d}V_{i,r_\ell(k)}^{1-q_\ell(k)}
\{1-V_{i,r_\ell(k)}\}^{q_\ell(k)},
\qquad
z_i\sim\operatorname{Categorical}(\phi_{i1},\ldots,\phi_{iK}).
\label{eq:profile_probability}
\end{equation}
Each terminal probability therefore requires only $\log_2K$ branching decisions.

\subsection{Engagement-Associated Acquisition}
\label{subsec:acquisition}

Engagement combines activity count, mean duration, and engagement ratio multiplicatively, with missing or negative values set to zero. We log-transform and standardize the result:
\[
x^{\log}_{it}=\log(1+x^{\mathrm{raw}}_{it}),
\qquad
x_{it}=(x^{\log}_{it}-\bar{x}^{\log})/s_x,
\]
where the mean and standard deviation are taken over observed student-chapter pairs. 

For profile $k$ and topic $m$, acquisition is
\begin{equation}
f_{km}(x)=\sum_{\ell=1}^{L}B_{\ell}(x)\beta_{km\ell},
\qquad
\beta_{km1}=\beta^{(0)}_{km}\sim N(a_k+c_m,\sigma_{\beta0}^2),
\label{eq:acquisition}
\end{equation}
where $a_k$ is the ordered profile baseline and $c_m$ is a fixed, centered topic anchor. 
Ordered baseline locations stabilize profile labels: fixed prior anchors are equally spaced from $1.20$ to $-1.00$, $a_1$ is stochastic, and $a_{k+1}=a_k-g^a_k$ with $g^a_k>0.02$. Hence $a_1>\cdots>a_K$. This ordering applies only to baseline locations, not transfer persistence. Prior distributions and anchor choices are detailed in Appendix~\ref{app:implementation}.
We use $L=8$ cubic basis functions, boundary knots at the observed engagement range, and internal knots at empirical quantiles. 

Positive, decreasing adjacent increments regularize acquisition toward monotone diminishing returns. For total gain $G_{km}>0$, set $u_{km1}=1$ and
\[
u_{km,\ell+1}=u_{km\ell}\rho_{km\ell},
\qquad \rho_{km\ell}\sim\operatorname{Beta}(4,2),
\]
then, for $\ell=1,\ldots,L-1$,
\begin{equation}
\Delta_{km\ell}
=G_{km}\frac{u_{km\ell}}{\sum_{r=1}^{L-1}u_{kmr}},
\qquad
\beta_{km,\ell+1}=\beta_{km\ell}+\Delta_{km\ell}.
\label{eq:delta}
\end{equation}
Thus, $\Delta_{km\ell}>0$ and $\sum_{\ell=1}^{L-1}\Delta_{km\ell}=G_{km}$. The total gain summarizes overall coefficient-level improvement, while $\rho_{km\ell}$ controls how quickly it accumulates. 

\begin{proposition}[Monotone acquisition]
\label{prop:monotone_acquisition}
Suppose the B-spline basis is normalized and nonnegative and
$\beta_{km1}\leq\cdots\leq\beta_{kmL}$. Then $f_{km}(x)$ is nondecreasing over the spline domain.
\end{proposition}

The positive-increment construction therefore guarantees monotone acquisition. Decreasing increments further regularize the curves toward diminishing returns but do not imply concavity for an arbitrary knot sequence. The proof is given in Appendix~\ref{app:model_properties}.

\subsection{Curriculum-Mediated Dynamic Transfer}
\label{subsec:dynamic_transfer}

The expert-designed network \citep{CourseKata2023} described in Section~\ref{sec:motivation} supplies pedagogical similarities $S_{tc}\geq0$, with $S_{tc}=0$ for $c\geq t$. We normalize within each receiving chapter:
\begin{equation}
W_{tc}=\frac{S_{tc}}{\sum_{h<t}S_{th}},\qquad c<t,
\label{eq:W}
\end{equation}
when the denominator is positive, and set $W_{tc}=0$ otherwise. Thus, $\mathbf W$ is nonnegative, strictly lower triangular, and row-normalized for chapters with earlier connections. Rows index receiving chapters and columns source chapters; the computational notation is $\omega_{ct}=W_{tc}$.
Conditional on $z_i=k$,
\begin{equation}
\theta_{it}\sim N(\mu_{it},\sigma_\epsilon^2),
\label{eq:theta_dist}
\end{equation}
where
\begin{equation}
\mu_{i1}=f_{k,m(1)}(x_{i1})+\delta_1
\label{eq:mu_ch1}
\end{equation}
and, for $t=2,\ldots,T$,
\begin{equation}
\mu_{it}=f_{k,m(t)}(x_{it})
+\lambda_k\sum_{c<t}W_{tc}\theta_{ic}+\delta_t.
\label{eq:dynamic}
\end{equation}
The terms represent current acquisition, curriculum-mediated dependence, and residual chapter effects. The centered effects satisfy $\delta_t=\delta^{\mathrm{raw}}_t-T^{-1}\sum_{h=1}^{T}\delta^{\mathrm{raw}}_h$ and $\sum_t\delta_t=0$.

\paragraph{Transfer persistence and propagation.}
The parameter $\lambda_k\in(0,1)$ controls profile-specific persistence. Independent logit-scale priors use weak anchors equally spaced from $0.56$ to $0.44$:
\begin{equation}
\eta_k=\operatorname{logit}(\lambda_k),
\qquad \eta_k\sim N(\eta^{\star}_k,0.75^2).
\label{eq:lambda_prior}
\end{equation}
Higher baseline response propensity therefore need not imply stronger transfer.

For profile $k$, define $\mathbf H_k=\lambda_k\mathbf W$, let $\boldsymbol\theta_i=(\theta_{i1},\ldots,\theta_{iT})^\top$, and let $\mathbf u_i$ collect $f_{k,m(t)}(x_{it})+\delta_t$ over chapters. Then $\boldsymbol\theta_i=\mathbf H_k\boldsymbol\theta_i+\mathbf u_i+\boldsymbol\epsilon_i$.

\begin{theorem}[Finite curriculum-path representation]
\label{thm:path}
For each learner profile $k$, $\mathbf H_k^r=\mathbf 0$ for $r\geq T$, and
\begin{equation}
\boldsymbol\theta_i
=(\mathbf I-\mathbf H_k)^{-1}(\mathbf u_i+\boldsymbol\epsilon_i)
=\sum_{r=0}^{T-1}\mathbf H_k^r(\mathbf u_i+\boldsymbol\epsilon_i).
\label{eq:path_expansion}
\end{equation}
Moreover, for $c<t$,
\begin{equation}
\tau_{k,tc}
=\left[(\mathbf I-\lambda_k\mathbf W)^{-1}-\mathbf I\right]_{tc}
=\sum_{r=1}^{T-1}\lambda_k^r[\mathbf W^r]_{tc},
\label{eq:tau_matrix}
\end{equation}
which sums the contributions of all directed curriculum paths from chapter $c$ to chapter $t$.
\end{theorem}

The $r=1$ term in $\tau_{k,tc}$ is direct propagation; terms with $r\geq2$ pass through intermediate chapters. Effective curriculum cohesion,
\begin{equation}
C_k=\frac{2}{T(T-1)}\sum_{t>c}\tau_{k,tc},
\label{eq:cohesion}
\end{equation}
averages this propagation over ordered chapter pairs. Both quantities are model-based sensitivity summaries, not causal effects. The proof and additional propagation properties appear in Appendix~\ref{app:model_properties}; remaining prior specifications appear in Appendix~\ref{app:implementation}.

\subsection{Posterior Inference and Educational Interpretation}
\label{subsec:inference}

We fit the model by MCMC in NIMBLE with different $K$. Convergence is assessed using multiple chains, trace plots, effective sample sizes, and potential scale reduction diagnostics. Model choice combines WAIC with posterior predictive adequacy, effective profile sizes, classification certainty, and substantive stability. Predictive checks use overall accuracy and chapter-level variation in accuracy. Appendix~\ref{app:implementation} provides the monitored quantities, sampler, and WAIC calculation.
Because transfer is unordered and real-data profile labels are unknown, posterior summaries are aligned draw by draw by the average terminal spline coefficient across topics. Aligned Profile 1 has the highest terminal acquisition level in each draw. This alignment is used only for summarization and visualization; it does not constrain transfer during fitting.
The case-study summaries are $f_{km}$ and $G_{km}$ for acquisition, $\lambda_k$, $\tau_{k,tc}$, and $C_k$ for transfer, and profile probabilities and $\boldsymbol\gamma_r$ for learner heterogeneity. 

\section{Simulation Validation}
\label{sec:simulation}

We validated \textsc{NAFI} using 200 Monte Carlo replications calibrated to the CourseKata sample size, curriculum length, item density, topic structure, and engagement measurement. Model comparison for $K\in\{2,4,8\}$ favored the true four-profile specification over under-extraction and removal of curriculum transfer, while eight profiles provided no meaningful predictive gain. Under the correctly specified model, aligned learner-profile classification accuracy averaged 0.955; transfer parameters were recovered with average bias $-0.001$ and RMSE $0.080$; and engagement-acquisition curves achieved mean curve-level RMSE $0.127$ and pointwise coverage $0.964$, even though the generating curves were exponential rather than B-splines. These results support recovery of the three mechanisms motivating the empirical analysis while indicating greater uncertainty for covariate effects and some profile-specific curves. The full simulation design, evaluation metrics, tables, figures, and detailed results are provided in Appendix~\ref{app:simulation}.

\section{Learning Acquisition and Curriculum Transfer in CourseKata}
\label{sec:application}

We now return to the three empirical features that motivated the model in Figure~\ref{fig:coursekata}: variation in performance and engagement across chapters, nonlocal connections in the curriculum, and heterogeneous student trajectories. The analysis below addresses the corresponding educational questions in the same order. We first examine engagement-associated acquisition across topics (Q1), then curriculum-mediated transfer across chapters (Q2), and finally differences in these learning patterns across learner profiles and presurvey characteristics (Q3). All estimates remain associational and are interpreted as model-based summaries of the observed longitudinal response patterns.

Before addressing these questions, we compared models with $K\in\{2,4,8\}$ and selected $K=4$ as the best balance of predictive fit, classification quality, and substantive interpretability. Although $K=8$ yielded slightly lower WAIC, it produced an empty profile and substantially weaker classification stability; $K=2$ had substantially worse predictive fit. Ablation checks further supported retaining both learner profiles and curriculum-mediated transfer. The full model-comparison and ablation results are reported in Appendix~\ref{app:coursekata_model_selection}.

\subsection{Empirical Answers to the Motivating Questions}

\subsubsection{Engagement-Associated Acquisition Across Topics (Q1)}
\label{subsec:real_acquisition}

\begin{figure}[!htbp]
    \centering
\captionsetup[subfloat]{font=scriptsize}
    \subfloat[Estimated engagement-acquisition curves.]{
        \includegraphics[
            width=0.45\textwidth,
            height=0.25\textheight,
            keepaspectratio
        ]{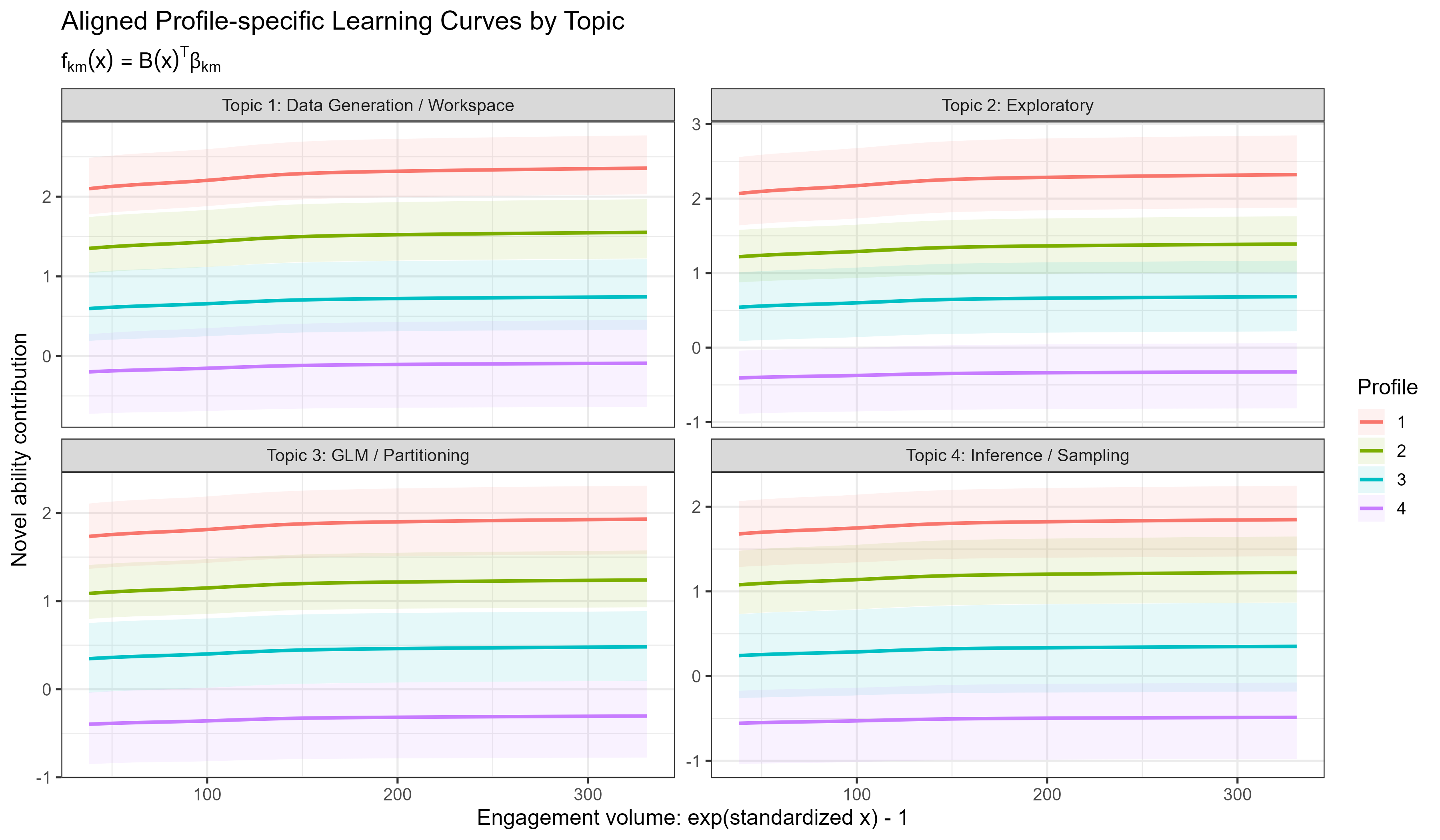}
        \label{fig:LearningCurve}
    }
    \subfloat[Estimated engagement-associated learning gains.]{
        \includegraphics[
            width=0.45\textwidth,
            height=0.25\textheight,
            keepaspectratio
        ]{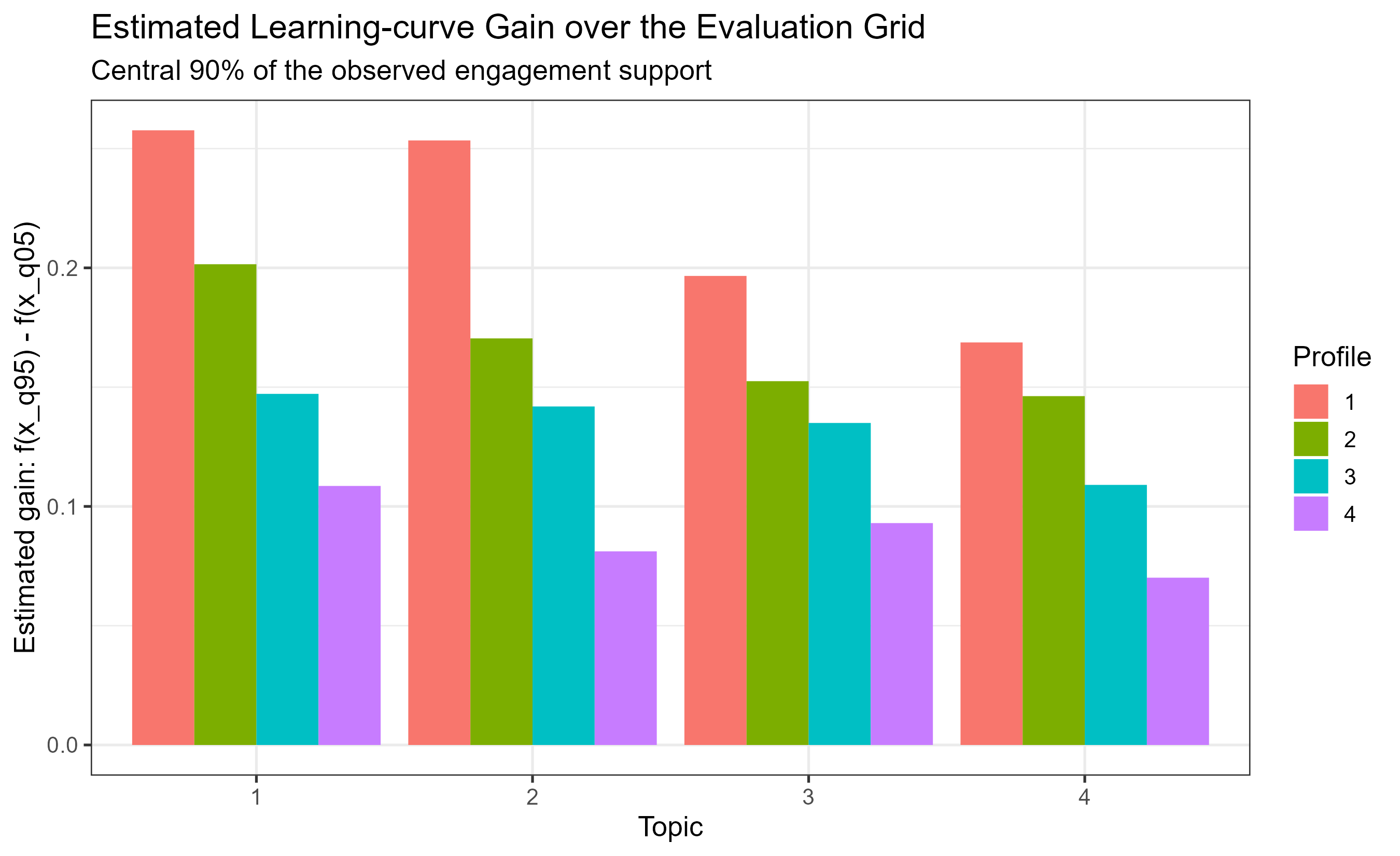}
        \label{fig:LearningGain}
    }
    \caption{Estimated engagement-acquisition curves and engagement-associated learning gains by learner profile and pedagogical topic.}
    \label{fig:LearningCurveGainCombined}
\end{figure}

The chapter-level variation in performance and engagement in Figures~\ref{fig:coursekata}(a)--(b) motivates our first question: how is engagement with current material associated with new knowledge acquisition? In the selected model, standardized engagement $x_{it}$ enters the acquisition component through a monotone total-gain spline,
\(
f_{km}(x)=\sum_{\ell=1}^{L}B_{\ell}(x)\beta_{km\ell},
\,
\sum_{\ell=1}^{L-1}\Delta_{km\ell}=G_{km}.
\)
Thus, $G_{km}$ summarizes the total engagement-related increase for profile $k$ and topic $m$, while the normalized increments determine how the increase is distributed over the engagement range.
Figure~\ref{fig:LearningCurve} shows the estimated acquisition curves by topic and profile.
The curves reveal substantial heterogeneity in both baseline acquisition levels and engagement-related gains.
Across topics, Profile 1 consistently had the highest acquisition functions, followed by Profiles 2, 3, and 4.
This ordering foreshadows the learner-profile differences addressed under Q3: learners with stronger prior preparation, lower anxiety, and higher engagement also showed higher estimated acquisition levels.

We summarize gains as $f_{km}(x_{0.95})-f_{km}(x_{0.05})$ in Figure~\ref{fig:LearningGain}, where $x_{0.05}$ and $x_{0.95}$ denote the lower and upper bounds of the central 90\% observed engagement range.
The gains were positive across profiles and topics, indicating that greater engagement was generally associated with higher latent acquisition. However, the magnitude of this gain varied substantially across profiles.
Profile 1 showed the largest engagement-related gains, whereas Profile 4 showed the weakest gains.
Thus, the answer to Q1 is not a single common learning curve. 
Greater engagement was generally associated with higher latent acquisition, 
but both the baseline level of acquisition and the size of the engagement-related gain differed across topics and learner profiles. 
In particular, students with weaker prior preparation and lower baseline performance showed weaker associations between additional engagement and acquisition, suggesting that time-on-task alone may not yield comparable learning benefits for all learners.

\subsubsection{Curriculum-Mediated Transfer Across Chapters (Q2)}
\label{subsec:real_transfer}

The nonadjacent pedagogical links in Figure~\ref{fig:coursekata}(c) motivate our second question: how does knowledge developed earlier contribute to learning later in the curriculum? The dynamic component represents this contribution through
\[
\lambda_{z_i}\sum_{c<t}\omega_{ct}\theta_{ic}.
\]
In the selected $K=4$ model, the posterior means of $\lambda_k$ were positive across all profiles.
Stronger performance in prior chapters was therefore associated with higher latent response propensity in later chapters connected through the curriculum network, including nonadjacent chapters.
Although the posterior means differ numerically, the overlapping intervals indicate that the data provide stronger support for the existence of positive curriculum-linked dependence than for large profile-specific differences in transfer strength.

\begin{figure}[!htbp]
    \centering
    \includegraphics[scale=0.4]{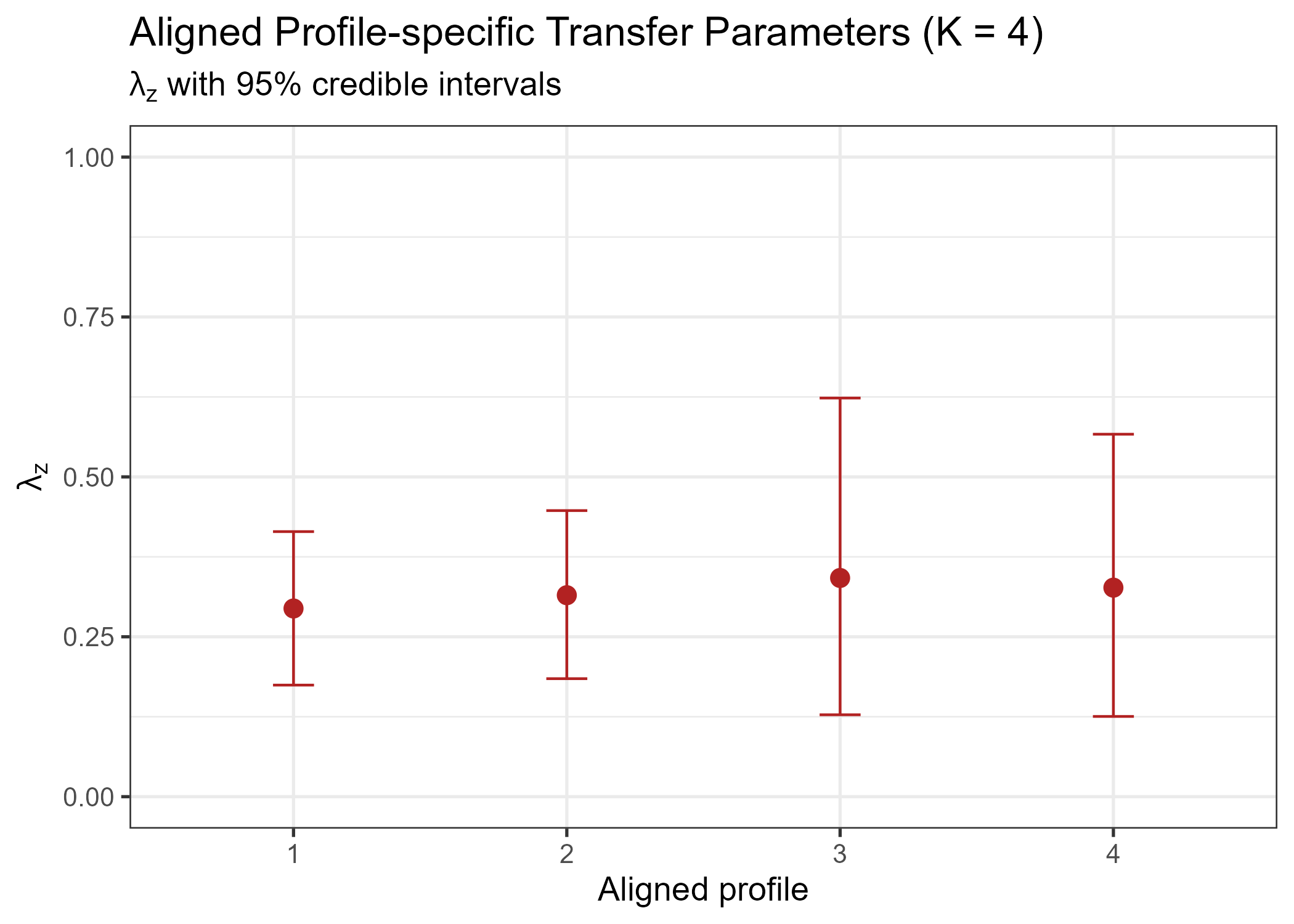}
    \caption{Posterior means and credible intervals for profile-specific curriculum-transfer parameters.}
    \label{fig:transfer}
\end{figure}

\begin{figure}[!htbp]
    \centering
    \captionsetup[subfloat]{font=scriptsize}
    \subfloat[Estimated centered chapter effects.]{
        \includegraphics[width=0.4\textwidth]{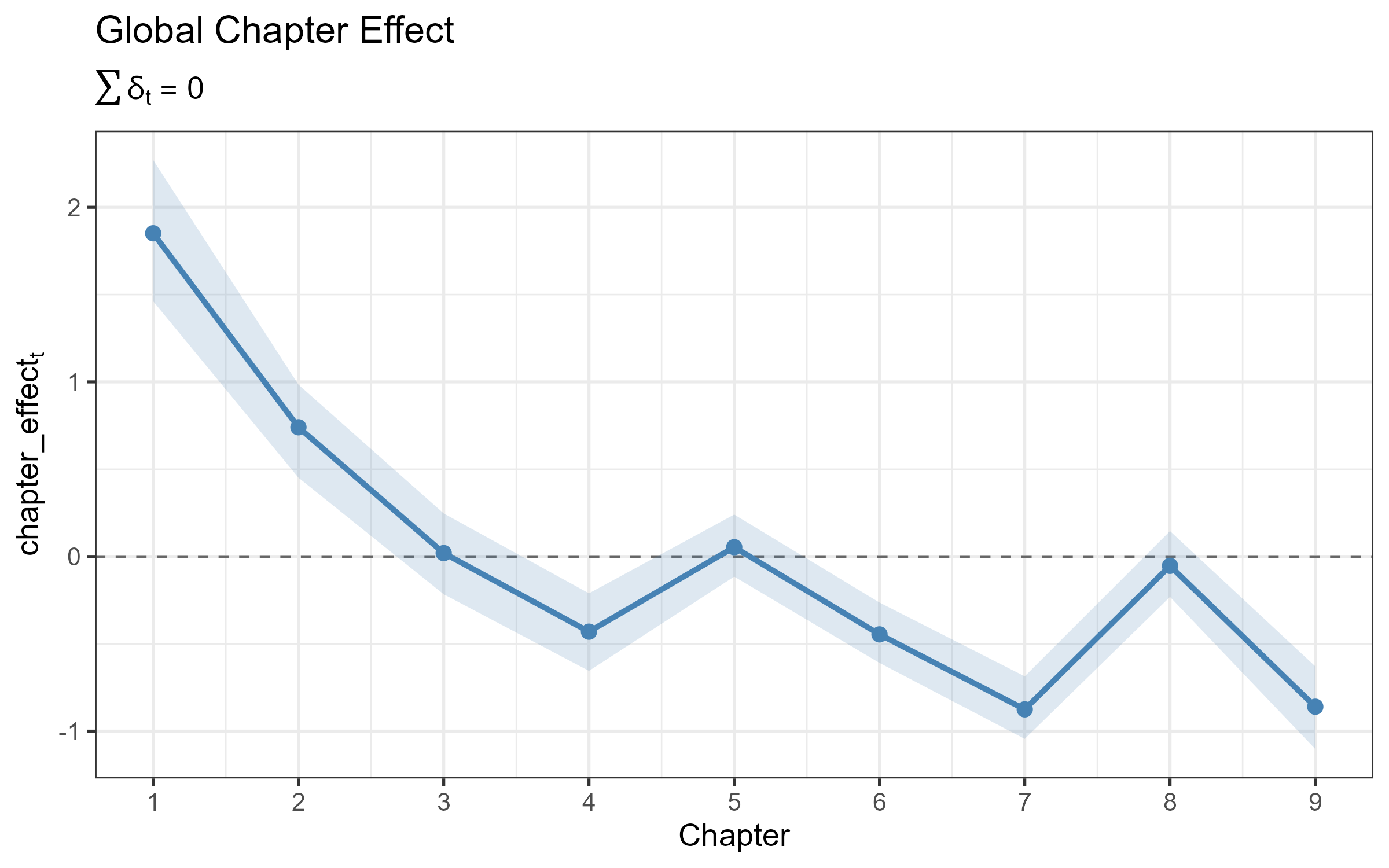}
        \label{fig:chapterEffects}
    }
    \hfill
    \subfloat[Estimated chapter effects in relation to observed chapter-level accuracy.]{
        \includegraphics[width=0.4\textwidth]{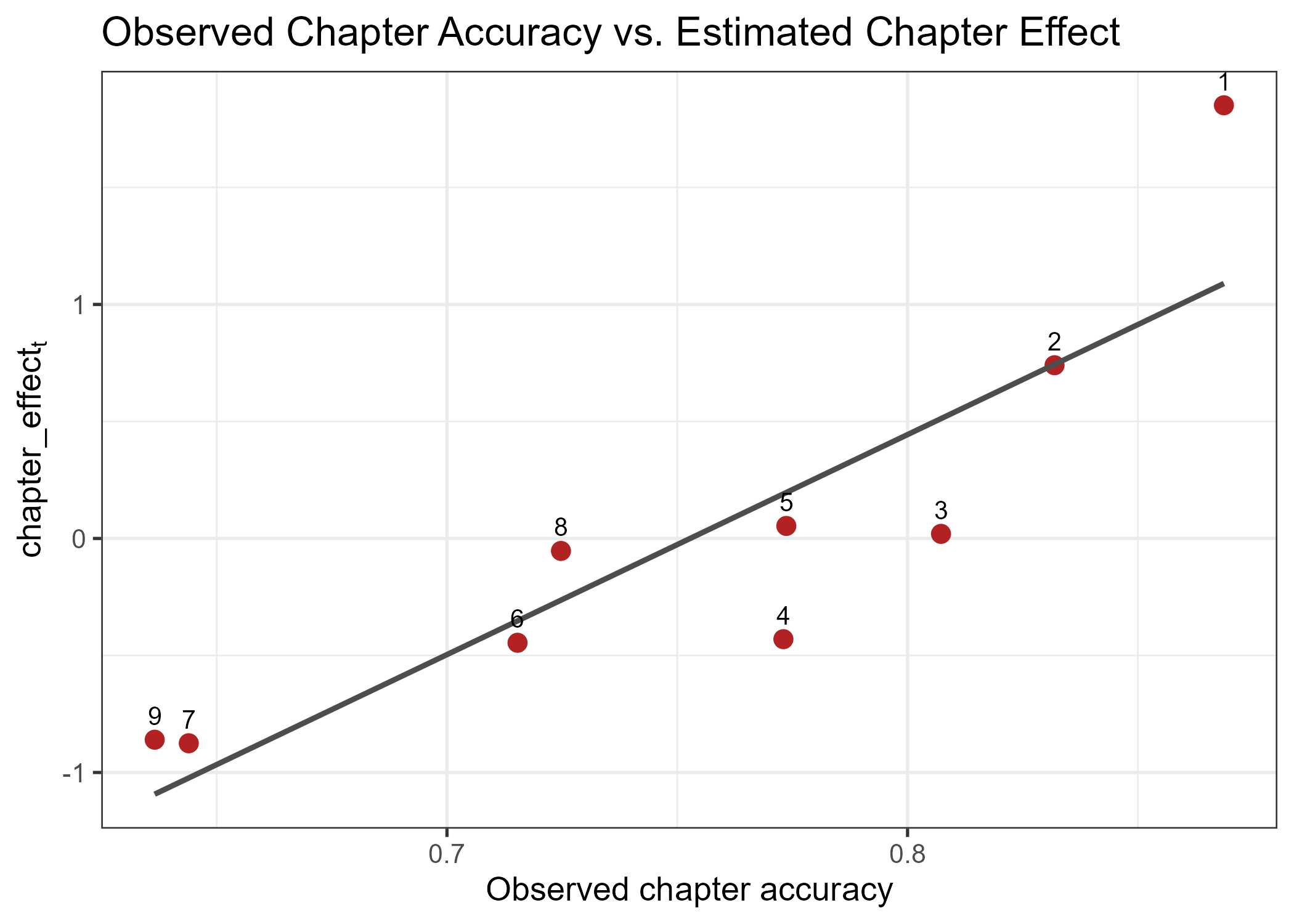}
        \label{fig:chapterAcc}
    }
    \caption{Estimated chapter effects and their relation to observed chapter-level accuracy.}
    \label{fig:chapterEffectsCombined}
\end{figure}

We also examined the centered chapter effects $\delta_t$, which represent residual chapter-level shifts after accounting for engagement-associated acquisition, curriculum-mediated transfer, learner profiles, and item difficulties.
Figure~\ref{fig:chapterEffects} shows that Chapters 1 and 2 had positive residual effects, whereas Chapters 7 and 9 had the most negative residual effects.
These patterns echo the observed chapter-level performance differences in Figure~\ref{fig:coursekata}(a); the direct comparison with observed accuracy is shown in Figure~\ref{fig:chapterAcc}.
Negative residual effects identify chapters where students performed worse than expected given their engagement, profile membership, prior curriculum performance, and item difficulty.
Chapters 7 and 9 therefore emerge as potential curriculum bottlenecks whose prerequisite dependencies, explanations, practice opportunities, and assessment alignment may warrant closer inspection.

\subsubsection{Learner Profiles and Presurvey Characteristics (Q3)}
\label{subsec:real_profiles}

Finally, the heterogeneous student-by-chapter trajectories in Figure~\ref{fig:coursekata}(d) motivate our third question: do students differ in how they acquire new knowledge and build on earlier learning, and are these differences related to presurvey characteristics? Under the selected $K=4$ model, posterior modal classification identified four learner profiles with distinct combinations of performance, engagement, prior preparation, affective characteristics, and learning beliefs.
The largest profile contained 105 students, whereas the smallest contained 12 students.
The average maximum posterior profile probability was 0.800, indicating reasonably well-identified but uncertain profile assignments.

Table~\ref{tab:character} summarizes the student-level characteristics of the four profiles.
Profile 1 comprised well-prepared, high-engagement learners with the highest observed accuracy, strongest prior mathematics preparation, lowest mathematics anxiety, and highest engagement.
Profile 2, the largest subgroup, combined moderate-to-high accuracy and relatively high engagement with elevated mathematics anxiety and memorization-oriented learning beliefs.
Profile 3 showed lower observed accuracy and the highest mathematics anxiety, together with relatively high memorization-oriented beliefs.
Profile 4 was a small at-risk subgroup with the lowest observed accuracy, weakest prior mathematics preparation, and lowest engagement.
Together with the profile-specific acquisition curves under Q1, these differences show that students followed distinct learning patterns rather than simply occupying different points on a single performance scale.

\begin{table}[!htbp]
\centering
\caption{Characteristics of the four learner profiles.}
\label{tab:character}
\resizebox{\textwidth}{!}{
\begin{tabular}{lrrrrrrr}
\toprule
Profile & Size & Observed accuracy & Engagement & Math anxiety & Prior math & Memorization belief & Coding experience \\
\hline
1 &  57 & 0.880 & 159.984 & 0.333 & 0.772 & 0.719 & 0.667 \\
2 &  105 & 0.764 & 135.984 & 0.581 & 0.438 & 0.733 & 0.638 \\
3 &  24 & 0.602 & 134.542 & 0.708 & 0.458 & 0.792 & 0.583 \\
4 &  12 & 0.403 & 124.253 & 0.500 & 0.250 & 0.833 & 0.583\\
\bottomrule
\end{tabular}}
\end{table}

\begin{figure}[!htbp]
    \centering
    \subfloat[Observed characteristics by learner profile.]{
        \includegraphics[width=0.45\textwidth]{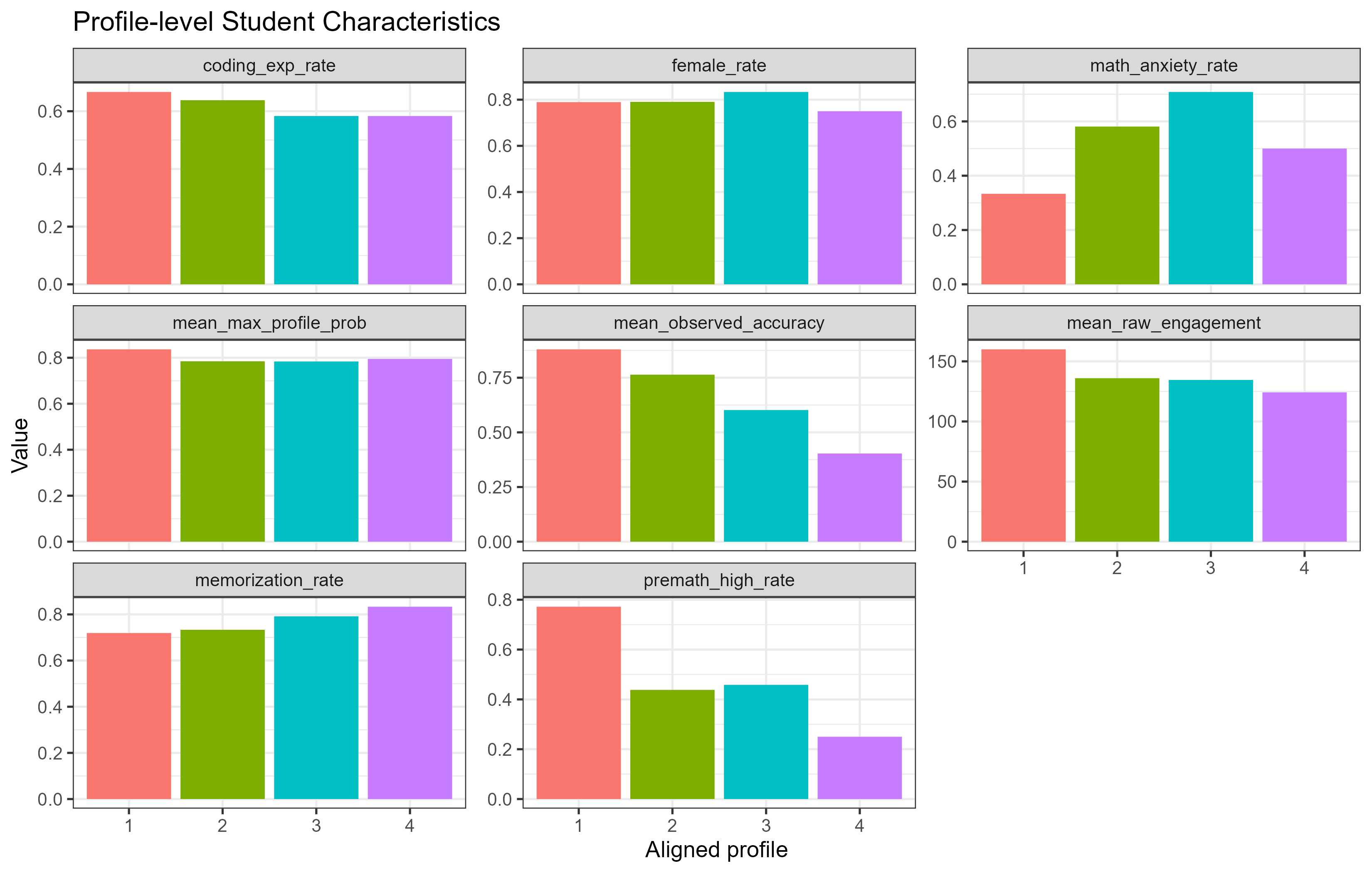}
        \label{fig:ProfileChar}
    }
    \subfloat[Estimated presurvey covariate effects.]{
        \includegraphics[width=0.45\textwidth]{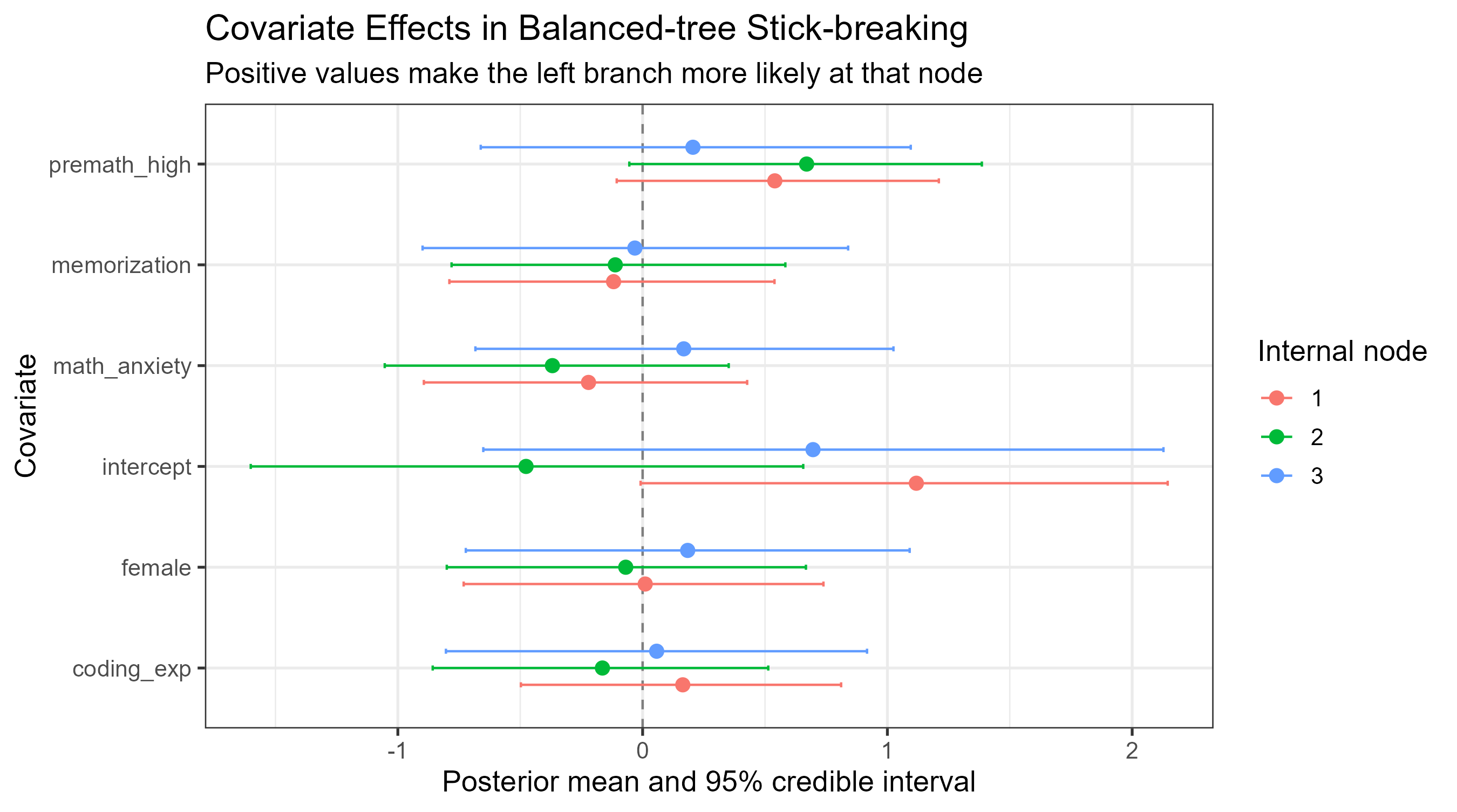}
        \label{fig:covariate}
    }
    \caption{Learner-profile characteristics and presurvey covariate effects in the balanced-tree stick-breaking profile model.}
    \label{fig:profileCovariateCombined}
\end{figure}

Figure~\ref{fig:ProfileChar} visualizes these profile-level differences.
Although the profiles also differed somewhat in observed gender composition and coding experience, the covariate effect estimates below provide limited evidence that these variables systematically drove profile membership.

Covariate effects in the balanced-tree stick-breaking model are reported in Figure~\ref{fig:covariate}.
These coefficients characterize how student background variables are associated with profile membership through internal tree-branching probabilities, with positive values making assignment to the left branch more likely at a given node.
Overall, the effects were estimated with considerable uncertainty, and most credible intervals overlapped zero.
Prior mathematics preparation showed the most consistent positive association with left-branch assignment, while mathematics anxiety tended to have negative effects at some nodes.
Gender, memorization beliefs, and coding experience showed little evidence of systematic associations with profile membership.
Thus, the descriptive differences across profiles should be interpreted primarily as profile-level patterns rather than as sharply identified covariate-driven assignment mechanisms.



\subsection{Implications for Online Learning Practice}
The empirical answers to Q1--Q3 lead to three practical implications.
First, the acquisition results show that engagement was generally
associated with higher acquisition, but the size of the gain varied
across profiles. Simply increasing time-on-task or completed activities
may therefore be insufficient, especially for at-risk learners.
Online platforms may need to scaffold engagement through adaptive
hints, worked examples, targeted review, and metacognitive prompts.
Instructors should consider not only how much students engage but
also whether their engagement translates into meaningful knowledge
acquisition. Monitoring acquisition trajectories alongside engagement
could help identify students who remain active but make limited progress.
Second, the positive curriculum-mediated transfer parameters indicate
that prior chapter performance is informative for later learning within
the curriculum network. This supports curriculum-aware adaptive
sequencing: when a student struggles in an earlier chapter that
supports later topics, the platform may provide targeted review
before the student proceeds. The negative residual effects for
Chapters 7 and 9 identify priorities for instructional redesign,
including clearer explanations, additional formative assessment,
prerequisite reminders, and more gradual transitions between concepts.
More broadly, estimated transfer pathways could help instructors
identify foundational chapters that deserve additional attention
because performance in these chapters is associated with learning
across multiple subsequent topics.
Third, the learner profiles suggest moving beyond one-size-fits-all
interventions. Profile 2 students performed relatively well but
began the course with higher mathematics anxiety, suggesting a need
for affective support and feedback that promotes confidence and
conceptual understanding. These students may not be identified by
accuracy-based early warning systems, despite their initially high
mathematics anxiety and memorization-oriented beliefs that suggest
potential benefits from affective and metacognitive support.
Profile 4 students combined low prior preparation, low engagement,
and low performance, suggesting a need for early warning systems,
prerequisite review, and intensive remediation. Together, these
profiles illustrate how longitudinal learning patterns and presurvey
characteristics can complement conventional performance measures
in identifying students who may benefit from additional support.
They also provide a basis for teaching experiments that compare
differentiated instructional strategies across learner profiles
and evaluate their effects on knowledge acquisition and transfer.

\section{Conclusion and Discussion}
\label{sec:discussion}

We developed \textsc{NAFI} to study acquisition, curriculum-mediated transfer, and learner heterogeneity in longitudinal embedded-assessment data. The model combines Rasch-type measurement, monotone acquisition splines, an expert-designed curriculum network, and covariate-dependent learner profiles. Simulations calibrated to CourseKata support recovery of these components, including acquisition curves under functional misspecification.
The CourseKata analysis produced three main findings. First, engagement was generally associated with higher latent response propensity, but baseline acquisition and engagement-associated gains differed across profiles and topics. The lowest-performing profile showed weaker gains, suggesting that additional time-on-task may need to be accompanied by scaffolding, feedback, and prerequisite support. Second, transfer parameters were positive across profiles, including through nonadjacent curriculum connections. Their overlapping intervals provide stronger evidence of general curriculum-linked dependence than of sharply different transfer strengths. Negative residual effects in Chapters 7 and 9 identified possible curriculum bottlenecks. Third, the four profiles differed in accuracy, engagement, prior preparation, mathematics anxiety, and learning beliefs. In particular, relatively strong performance could coexist with elevated anxiety, while the smallest profile combined low preparation, engagement, and performance. The uncertainty in covariate effects and profile assignments argues against treating these profiles as fixed student types.
These findings suggest three practical directions: scaffold engagement rather than simply increase activity, use curriculum connections to target prerequisite review, and look beyond accuracy when identifying students who may need support. 

Several directions offer opportunities for future research.
One promising extension is to incorporate richer measures of
engagement quality to better understand how different learning
behaviors relate to knowledge acquisition and transfer.
Another direction is to develop data-driven approaches for
learning and refining curriculum networks, allowing the
relationships among chapters to be informed by both pedagogical
knowledge and student response patterns. It would also be
interesting to extend the framework to finer curricular
scales, such as individual pages and learning activities,
to characterize more detailed learning processes.
Finally, integrating the proposed framework with randomized
or quasi-experimental designs could provide new opportunities
to evaluate how instructional interventions influence
knowledge acquisition, transfer, and heterogeneous
learning trajectories.

\section*{Data and Code Availability Statement}
The CourseKata data used in this study are publicly available by request for research purposes. 
The R code for implementing the proposed model is available on Github \hyperlink{https://github.com/mazh1993/NAFI.git}{github/NAFI}.

\bibliographystyle{agsm}

\bibliography{Bibliography-MM-MC}

@article{disessa1993epistemology,
  author  = {diSessa, Andrea A.},
  title   = {Toward an Epistemology of Physics},
  journal = {Cognition and Instruction},
  year    = {1993},
  volume  = {10},
  number  = {2-3},
  pages   = {105--225},
  doi     = {10.1080/07370008.1985.9649008}
}

@article{disessa2014causal,
  author  = {diSessa, Andrea A.},
  title   = {The Construction of Causal Schemes:
             Learning Mechanisms at the Knowledge Level},
  journal = {Cognitive Science},
  year    = {2014},
  volume  = {38},
  number  = {5},
  pages   = {795--850},
  doi     = {10.1111/cogs.12131}
}

@book{linn2011science,
  author    = {Linn, Marcia C. and Eylon, Bat-Sheva},
  title     = {Science Learning and Instruction:
               Taking Advantage of Technology to Promote Knowledge Integration},
  publisher = {Routledge},
  year      = {2011},
  doi       = {10.4324/9780203806524}
}

@article{bransford1999transfer,
  author  = {Bransford, John D. and Schwartz, Daniel L.},
  title   = {Rethinking Transfer:
             A Simple Proposal with Multiple Implications},
  journal = {Review of Research in Education},
  year    = {1999},
  volume  = {24},
  pages   = {61--100},
  doi     = {10.3102/0091732X024001061}
}

@article{rittlejohnson2001iterative,
  author  = {Rittle-Johnson, Bethany and Siegler, Robert S.
             and Alibali, Martha Wagner},
  title   = {Developing Conceptual Understanding and Procedural
             Skill in Mathematics: An Iterative Process},
  journal = {Journal of Educational Psychology},
  year    = {2001},
  volume  = {93},
  number  = {2},
  pages   = {346--362},
  doi     = {10.1037/0022-0663.93.2.346}
}

@article{eccles2020situated,
  author  = {Eccles, Jacquelynne S. and Wigfield, Allan},
  title   = {From Expectancy-Value Theory to Situated
             Expectancy-Value Theory: A Developmental, Social
             Cognitive, and Sociocultural Perspective on Motivation},
  journal = {Contemporary Educational Psychology},
  year    = {2020},
  volume  = {61},
  pages   = {101859},
  doi     = {10.1016/j.cedpsych.2020.101859}
}

@article{pekrun2006control,
  author  = {Pekrun, Reinhard},
  title   = {The Control-Value Theory of Achievement Emotions:
             Assumptions, Corollaries, and Implications for
             Educational Research and Practice},
  journal = {Educational Psychology Review},
  year    = {2006},
  volume  = {18},
  number  = {4},
  pages   = {315--341},
  doi     = {10.1007/s10648-006-9029-9}
}

@article{ashcraft2007working,
  author  = {Ashcraft, Mark H. and Krause, Jeremy A.},
  title   = {Working Memory, Math Performance, and Math Anxiety},
  journal = {Psychonomic Bulletin \& Review},
  year    = {2007},
  volume  = {14},
  number  = {2},
  pages   = {243--248},
  doi     = {10.3758/BF03194059}
}

@article{zimmerman2002selfregulated,
  author  = {Zimmerman, Barry J.},
  title   = {Becoming a Self-Regulated Learner: An Overview},
  journal = {Theory Into Practice},
  year    = {2002},
  volume  = {41},
  number  = {2},
  pages   = {64--70},
  doi     = {10.1207/s15430421tip4102_2}
}

@article{tamano2023dynamical,
  author  = {Tamano, Hiroshi and Mochihashi, Daichi},
  title   = {Dynamical Non-compensatory Multidimensional {IRT} Model Using Variational Approximation},
  journal = {Psychometrika},
  volume  = {88},
  number  = {2},
  pages   = {487--526},
  year    = {2023},
  doi     = {10.1007/s11336-023-09903-y}
}

@article{black2009formative,
  author  = {Black, Paul and Wiliam, Dylan},
  title   = {Developing the Theory of Formative Assessment},
  journal = {Educational Assessment, Evaluation and Accountability},
  year    = {2009},
  volume  = {21},
  number  = {1},
  pages   = {5--31},
  doi     = {10.1007/s11092-008-9068-5}
}

@article{chi2014icap,
  author  = {Chi, Michelene T. H. and Wylie, Ruth},
  title   = {The {ICAP} Framework: Linking Cognitive Engagement to Active Learning Outcomes},
  journal = {Educational Psychologist},
  year    = {2014},
  volume  = {49},
  number  = {4},
  pages   = {219--243},
  doi     = {10.1080/00461520.2014.965823}
}

@article{fries2021connections,
  author  = {Fries, Laura and Son, Ji Y. and Givvin, Karen B. and Stigler, James W.},
  title   = {Practicing Connections: A Framework to Guide Instructional Design for Developing Understanding in Complex Domains},
  journal = {Educational Psychology Review},
  year    = {2021},
  volume  = {33},
  number  = {2},
  pages   = {739--762},
  doi     = {10.1007/s10648-020-09561-x}
}

@article{tetzlaff2021personalized,
  author  = {Tetzlaff, Leonard and Schmiedek, Florian and Brod, Garvin},
  title   = {Developing Personalized Education: A Dynamic Framework},
  journal = {Educational Psychology Review},
  year    = {2021},
  volume  = {33},
  number  = {3},
  pages   = {863--882},
  doi     = {10.1007/s10648-020-09570-w}
}

@article{hart2016precision,
  author  = {Hart, Sara A.},
  title   = {Precision Education Initiative: Moving Toward Personalized Education},
  journal = {Mind, Brain, and Education},
  year    = {2016},
  volume  = {10},
  number  = {4},
  pages   = {209--211},
  doi     = {10.1111/mbe.12109}
}

@article{hoch2018interactive,
  author  = {Hoch, Stefan and Reinhold, Frank and Werner, Bernhard and Richter-Gebert, J{\"u}rgen and Reiss, Kristina},
  title   = {Design and Research Potential of Interactive Textbooks: The Case of Fractions},
  journal = {{ZDM} Mathematics Education},
  year    = {2018},
  volume  = {50},
  number  = {5},
  pages   = {839--848},
  doi     = {10.1007/s11858-018-0971-z}
}

@article{kovanovic2015time,
  author  = {Kovanovi{\'c}, Vitomir and Ga{\v{s}}evi{\'c}, Dragan and Dawson, Shane and Joksimovi{\'c}, Sre{\'c}ko and Baker, Ryan S. and Hatala, Marek},
  title   = {Does Time-on-Task Estimation Matter? Implications on Validity of Learning Analytics Findings},
  journal = {Journal of Learning Analytics},
  year    = {2015},
  volume  = {2},
  number  = {3},
  pages   = {81--110},
  doi     = {10.18608/jla.2015.23.6}
}

@article{wang2013dynamic,
  author  = {Wang, Xiaojing and Berger, James O. and Burdick, Donald S.},
  title   = {Bayesian Analysis of Dynamic Item Response Models in Educational Testing},
  journal = {The Annals of Applied Statistics},
  year    = {2013},
  volume  = {7},
  number  = {1},
  pages   = {126--153},
  doi     = {10.1214/12-AOAS608}
}

@article{jeon2016autoregressive,
  author  = {Jeon, Minjeong and Rabe-Hesketh, Sophia},
  title   = {An Autoregressive Growth Model for Longitudinal Item Analysis},
  journal = {Psychometrika},
  volume  = {81},
  number  = {3},
  pages   = {830--850},
  year    = {2016},
  doi     = {10.1007/s11336-015-9489-2}
}

@inproceedings{zheng2025linking,
  author    = {Zheng, Kristine and Brockbank, Erik and Schwartz, Shawn T. and Yeager, David S. and Bryan, Christopher J. and Dweck, Carol S. and Fan, Judith E.},
  title     = {Linking Student Psychological Orientation, Engagement, and Learning in College-Level Introductory Data Science},
  booktitle = {Proceedings of the Annual Meeting of the Cognitive Science Society},
  year      = {2025},
  volume    = {47},
  pages     = {4071--4078},
  url       = {https://escholarship.org/uc/item/4fb8m5c9}
}

@book{Rasch1960,
  author    = {Rasch, Georg},
  title     = {Probabilistic Models for Some Intelligence and Attainment Tests},
  publisher = {Danish Institute for Educational Research},
  address   = {Copenhagen},
  year      = {1960},
  note      = {Expanded edition, University of Chicago Press, 1980}
}

@article{anderson1982acquisition,
  author  = {Anderson, John R.},
  title   = {Acquisition of Cognitive Skill},
  journal = {Psychological Review},
  volume  = {89},
  number  = {4},
  pages   = {369--406},
  year    = {1982},
  doi     = {10.1037/0033-295X.89.4.369}
}

@article{meyer2008inference,
  author  = {Meyer, Mary C.},
  title   = {Inference Using Shape-Restricted Regression Splines},
  journal = {The Annals of Applied Statistics},
  volume  = {2},
  number  = {3},
  pages   = {1013--1033},
  year    = {2008},
  doi     = {10.1214/08-AOAS167}
}

@article{chipman1998bayesian,
  author  = {Chipman, Hugh A. and George, Edward I. and McCulloch, Robert E.},
  title   = {Bayesian {CART} Model Search},
  journal = {Journal of the American Statistical Association},
  volume  = {93},
  number  = {443},
  pages   = {935--948},
  year    = {1998},
  doi     = {10.1080/01621459.1998.10473750}
}

@article{rockova2020posterior,
  author  = {Ro{\v{c}}kov{\'a}, Veronika and van der Pas, St{\'e}phanie},
  title   = {Posterior Concentration for {B}ayesian Regression Trees and Forests},
  journal = {The Annals of Statistics},
  volume  = {48},
  number  = {4},
  pages   = {2108--2131},
  year    = {2020},
  doi     = {10.1214/19-AOS1879}
}

@article{horiguchi2022tree,
  author  = {Horiguchi, Akira and Chan, Cliburn and Ma, Li},
  title   = {A Tree Perspective on Stick-Breaking Models in Covariate-Dependent Mixtures (with Discussion)},
  journal = {Bayesian Analysis},
  volume  = {20},
  number  = {3},
  pages   = {1139--1230},
  year    = {2025},
  doi     = {10.1214/24-BA1462}
}

@article{polson2013bayesian,
  author  = {Polson, Nicholas G. and Scott, James G. and Windle, Jesse},
  title   = {{B}ayesian Inference for Logistic Models Using {P}{\'o}lya--{G}amma Latent Variables},
  journal = {Journal of the American Statistical Association},
  volume  = {108},
  number  = {504},
  pages   = {1339--1349},
  year    = {2013},
  doi     = {10.1080/01621459.2013.829001}
}

@misc{CourseKata2023,
  author = {{CourseKata}},
  title  = {{CourseKata} Statistics and Data Science},
  year   = {2023},
  url    = {https://coursekata.org},
  note   = {Online learning platform}
}

@article{sutter2026perceptions,
  author  = {Sutter, Claudia C. and Jackson, Matthew C. and Son, Ji Y.},
  title   = {High School Students' Perceptions of and Attitudes Toward {R} Programming in Statistics and Data Science},
  journal = {Journal of Statistics and Data Science Education},
  year    = {2026},
  pages   = {1--16},
  doi     = {10.1080/26939169.2026.2635599},
  note    = {Advance online publication}
}

@article{stigler2020better,
  author  = {Stigler, James W. and Son, Ji Y. and Givvin, Karen B. and Blake, Adam B. and Fries, Laura and Shaw, Stacy T. and Tucker, Mary C.},
  title   = {The Better Book Approach for Education Research and Development},
  journal = {Teachers College Record},
  volume  = {122},
  number  = {9},
  pages   = {1--32},
  year    = {2020},
  doi     = {10.1177/016146812012200913}
}

@article{wang2020longitudinal,
  author  = {Wang, Chun and Nydick, Steven W.},
  title   = {On Longitudinal Item Response Theory Models: A Didactic},
  journal = {Journal of Educational and Behavioral Statistics},
  volume  = {45},
  number  = {3},
  pages   = {339--368},
  year    = {2020},
  doi     = {10.3102/1076998619882026}
}

@article{corbett1995knowledge,
  author  = {Corbett, Albert T. and Anderson, John R.},
  title   = {Knowledge Tracing: Modeling the Acquisition of Procedural Knowledge},
  journal = {User Modeling and User-Adapted Interaction},
  volume  = {4},
  number  = {4},
  pages   = {253--278},
  year    = {1995},
  doi     = {10.1007/BF01099821}
}

@article{barnett2002when,
  author  = {Barnett, Susan M. and Ceci, Stephen J.},
  title   = {When and Where Do We Apply What We Learn? {A} Taxonomy for Far Transfer},
  journal = {Psychological Bulletin},
  volume  = {128},
  number  = {4},
  pages   = {612--637},
  year    = {2002},
  doi     = {10.1037/0033-2909.128.4.612}
}

@inproceedings{cen2006learning,
  author    = {Cen, Hao and Koedinger, Kenneth R. and Junker, Brian},
  title     = {Learning Factors Analysis -- A General Method for Cognitive Model Evaluation and Improvement},
  booktitle = {Intelligent Tutoring Systems: 8th International Conference, ITS 2006},
  series    = {Lecture Notes in Computer Science},
  volume    = {4053},
  pages     = {164--175},
  publisher = {Springer},
  address   = {Berlin, Heidelberg},
  year      = {2006},
  doi       = {10.1007/11774303_17}
}

@inproceedings{pavlik2009performance,
  author    = {Pavlik, Philip I., Jr. and Cen, Hao and Koedinger, Kenneth R.},
  title     = {Performance Factors Analysis -- A New Alternative to Knowledge Tracing},
  booktitle = {Proceedings of the 14th International Conference on Artificial Intelligence in Education},
  series    = {Frontiers in Artificial Intelligence and Applications},
  volume    = {200},
  pages     = {531--538},
  publisher = {IOS Press},
  address   = {Amsterdam},
  year      = {2009}
}

@inproceedings{piech2015deep,
  author    = {Piech, Chris and Bassen, Jonathan and Huang, Jonathan and Ganguli, Surya and Sahami, Mehran and Guibas, Leonidas J. and Sohl-Dickstein, Jascha},
  title     = {Deep Knowledge Tracing},
  booktitle = {Advances in Neural Information Processing Systems},
  volume    = {28},
  pages     = {505--513},
  year      = {2015}
}

@article{chen2018hidden,
  author  = {Chen, Yinghan and Culpepper, Steven Andrew and Wang, Shiyu and Douglas, Jeffrey},
  title   = {A Hidden {M}arkov Model for Learning Trajectories in Cognitive Diagnosis With Application to Spatial Rotation Skills},
  journal = {Applied Psychological Measurement},
  volume  = {42},
  number  = {1},
  pages   = {5--23},
  year    = {2018}
}

@article{wang2018tracking,
  author  = {Wang, Shiyu and Yang, Yan and Culpepper, Steven Andrew and Douglas, Jeffrey A.},
  title   = {Tracking Skill Acquisition With Cognitive Diagnosis Models: A Higher-Order, Hidden {M}arkov Model With Covariates},
  journal = {Journal of Educational and Behavioral Statistics},
  volume  = {43},
  number  = {1},
  pages   = {57--87},
  year    = {2018},
  doi     = {10.3102/1076998617719727}
}

@article{ren2011logistic,
  author  = {Ren, Lu and Du, Lan and Carin, Lawrence and Dunson, David B.},
  title   = {Logistic Stick-Breaking Process},
  journal = {Journal of Machine Learning Research},
  volume  = {12},
  pages   = {203--239},
  year    = {2011}
}

@article{abbakumov2019measuring,
  author  = {Abbakumov, Dmitry and Desmet, Piet and Van den Noortgate, Wim},
  title   = {Measuring Growth in Students' Proficiency in {MOOC}s: Two Component Dynamic Extensions for the {R}asch Model},
  journal = {Behavior Research Methods},
  volume  = {51},
  number  = {1},
  pages   = {332--341},
  year    = {2019},
  doi     = {10.3758/s13428-018-1129-1}
}

@article{wang2016secondorder,
  author  = {Wang, Chun and Kohli, Nidhi and Henn, Lisa},
  title   = {A Second-Order Longitudinal Model for Binary Outcomes: Item Response Theory Versus Structural Equation Modeling},
  journal = {Structural Equation Modeling: A Multidisciplinary Journal},
  volume  = {23},
  number  = {3},
  pages   = {455--465},
  year    = {2016},
  doi     = {10.1080/10705511.2015.1096744}
}

@article{saqr2023longitudinal,
  author  = {Saqr, Mohammed and L{\'o}pez-Pernas, Sonsoles and Helske, Satu and Hrastinski, Stefan},
  title   = {The Longitudinal Association Between Engagement and Achievement Varies by Time, Students' Profiles, and Achievement State: A Full Program Study},
  journal = {Computers \& Education},
  volume  = {199},
  pages   = {104787},
  year    = {2023},
  doi     = {10.1016/j.compedu.2023.104787}
}

@article{dunson2006bayesian,
  author  = {Dunson, David B.},
  title   = {Bayesian Dynamic Modeling of Latent Trait Distributions},
  journal = {Biostatistics},
  volume  = {7},
  number  = {4},
  pages   = {551--568},
  year    = {2006}
}

@article{hickendorff2009solution,
  author  = {Hickendorff, Marian and Heiser, Willem J. and van Putten, Cornelis M. and Verhelst, Norman D.},
  title   = {Solution Strategies and Achievement in {D}utch Complex Arithmetic: Latent Variable Modeling of Change},
  journal = {Psychometrika},
  volume  = {74},
  number  = {2},
  pages   = {331--350},
  year    = {2009},
  doi     = {10.1007/s11336-008-9074-z}
}

@article{cho2010latent,
  author  = {Cho, Sun-Joo and Cohen, Allan S. and Kim, Seock-Ho and Bottge, Brian},
  title   = {Latent Transition Analysis With a Mixture Item Response Theory Measurement Model},
  journal = {Applied Psychological Measurement},
  volume  = {34},
  number  = {7},
  pages   = {483--504},
  year    = {2010},
  doi     = {10.1177/0146621610362978}
}

@article{nylund2023ten,
  author  = {Nylund-Gibson, Karen and Garber, Adam C. and Carter, Delwin B. and Chan, Mei-ki and Arch, D. A. N. and Simon, O. and Whaling, K. and Tartt, Erica and Lawrie, S. I.},
  title   = {Ten Frequently Asked Questions About Latent Transition Analysis},
  journal = {Psychological Methods},
  volume  = {28},
  number  = {2},
  pages   = {284--300},
  year    = {2023},
  doi     = {10.1037/met0000486}
}

@article{fox2005multilevel,
  author  = {Fox, Jean-Paul},
  title   = {Multilevel {IRT} Using Dichotomous and Polytomous Response Data},
  journal = {British Journal of Mathematical and Statistical Psychology},
  volume  = {58},
  number  = {1},
  pages   = {145--172},
  year    = {2005}
}

@article{pya2015shape,
  author  = {Pya, Natalya and Wood, Simon N.},
  title   = {Shape Constrained Additive Models},
  journal = {Statistics and Computing},
  volume  = {25},
  number  = {3},
  pages   = {543--559},
  year    = {2015},
  doi     = {10.1007/s11222-013-9448-7}
}

@article{cai2016item,
  author  = {Cai, Li and Choi, Kilchan and Hansen, Mark and Harrell, Lauren},
  title   = {Item Response Theory},
  journal = {Annual Review of Statistics and Its Application},
  volume  = {3},
  pages   = {297--321},
  year    = {2016},
  doi     = {10.1146/annurev-statistics-041715-033702}
}

@book{lord1968statistical,
  author    = {Lord, Frederic M. and Novick, Melvin R.},
  title     = {Statistical Theories of Mental Test Scores},
  publisher = {Addison-Wesley},
  address   = {Reading, MA},
  year      = {1968}
}

@book{baker2001basics,
  author    = {Baker, Frank B.},
  title     = {The Basics of Item Response Theory},
  edition   = {2nd},
  publisher = {ERIC Clearinghouse on Assessment and Evaluation},
  address   = {College Park, MD},
  year      = {2001}
}

@article{pan2024bayesian,
  author  = {Pan, Tianyu and Shen, Weining and Davis-Stober, Clintin P. and Hu, Guanyu},
  title   = {A {B}ayesian Nonparametric Approach for Handling Item and Examinee Heterogeneity in Assessment Data},
  journal = {British Journal of Mathematical and Statistical Psychology},
  volume  = {77},
  number  = {1},
  pages   = {196--211},
  year    = {2024},
  doi     = {10.1111/bmsp.12322}
}

@article{hu2024nonparametric, 
  author  = {Hu, Guanyu and Ma, Zhihua and Paek, Insu},
  title   = {A Nonparametric {B}ayesian Item Response Modelling Approach for Clustering Items and Individuals Simultaneously},
  journal = {Journal of Nonparametric Statistics},
  year    = {2024},
  doi     = {10.1080/10485252.2024.2354841}
}

@article{taple2022impact,
  author  = {Taple, Bayley J. and Chapman, Robert and Schalet, Benjamin D. and Brower, Rylee and Griffith, James W.},
  title   = {The Impact of Education on Depression Assessment: Differential Item Functioning Analysis},
  journal = {Assessment},
  volume  = {29},
  number  = {2},
  pages   = {272--284},
  year    = {2022},
  doi     = {10.1177/1073191120971357}
}

@article{shah2024economics,
  author  = {Shah, Syed Sibghatullah},
  title   = {The Economics of Education: Evaluating the Impact of Digital Learning Platforms},
  journal = {Premier Journal of Computer Science},
  year    = {2024},
  doi     = {10.70389/PJCS.100001}
}

@article{liu2020bayesian,
  author  = {Liu, Yang and Wang, Xiaojing},
  title   = {{B}ayesian Nonparametric Monotone Regression of Dynamic Latent Traits in Item Response Theory Models},
  journal = {Journal of Educational and Behavioral Statistics},
  volume  = {45},
  number  = {3},
  pages   = {274--296},
  year    = {2020},
  doi     = {10.3102/1076998619887913}
}

@article{liu2025general,
  author  = {Liu, Zichu and Wang, Shiyu and Xiao, Houping and Zhang, Shumei and Qiu, Tao},
  title   = {A General Dynamic Learning Model Framework for Cognitive Diagnosis},
  journal = {British Journal of Mathematical and Statistical Psychology},
  volume  = {78},
  pages   = {856--888},
  year    = {2025},
  doi     = {10.1111/bmsp.12384}
}

@article{sun2025bayesian,
  author  = {Sun, Jingyu and Liu, Yang and Wang, Xiaojing and Chen, Ming-Hui},
  title   = {{B}ayesian Variable Selection in Dynamic Item Response Theory Models},
  journal = {Journal of Educational and Behavioral Statistics},
  volume  = {51},
  number  = {2},
  pages   = {255--280},
  year    = {2026},
  doi     = {10.3102/10769986251314527}
}

\newpage
\setcounter{page}{1}
\appendix
\begin{center}
{\LARGE\bf Supplementary Material}
\end{center}
\section{Prior Specifications and Computational Details}
\label{app:implementation}

\subsection{Priors and Anchor Values}
The uncentered item difficulties satisfy
\[
b^{\mathrm{raw}}_{jt}\sim N(0,\sigma_{b,t}^2),
\qquad \sigma_{b,t}\sim\operatorname{HalfNormal}(0,2^2).
\]
Tree coefficients have $\gamma_{r1}\sim N(0,1^2)$ and $\gamma_{rp}\sim N(0,0.5^2)$ for $p\geq2$, allowing flexible marginal profile sizes while regularizing covariate associations.

The baseline anchors $a^{\star}_1,\ldots,a^{\star}_K$ are equally spaced from $1.20$ to $-1.00$, covering approximately $0.27$ to $0.77$ on the probability scale before item and chapter adjustments. We use
\[
a_1\sim N(a^{\star}_1,0.35^2),
\qquad
g^a_k\sim N(a^{\star}_k-a^{\star}_{k+1},0.25^2)
\ \text{truncated below at }0.02.
\]
The fixed topic anchors are $(c_1,c_2,c_3,c_4)=(0.10,0.05,-0.05,-0.10)$, and $\sigma_{\beta0}\sim\operatorname{HalfNormal}(0,0.5^2)$. These small centered offsets weakly regularize curve baselines; they are not estimated topic effects.

The total-gain prior is
\[
G_{km}\sim N(G^{\star}_{km},\sigma_G^2)\ \text{truncated below at }0,
\qquad \sigma_G\sim\operatorname{HalfNormal}(0,1^2),
\]
where $G^{\star}_{km}=g^{\star}_kh^{\star}_m$. The profile anchors $g^{\star}_k$ are equally spaced from $0.95$ to $0.35$, and $(h^{\star}_1,h^{\star}_2,h^{\star}_3,h^{\star}_4)=(1.00,0.95,0.85,0.75)$. These anchors provide weakly regularized positive starting values consistent with monotone acquisition.

For the dynamic components,
\[
\sigma_\epsilon\sim\operatorname{HalfNormal}(0,0.50^2),
\qquad
\delta^{\mathrm{raw}}_t\sim N(0,\sigma_\delta^2),
\qquad
\sigma_\delta\sim\operatorname{HalfNormal}(0,0.35^2).
\]
Transfer anchors $\lambda^{\star}_k$ are equally spaced from $0.56$ to $0.44$, with $\eta^{\star}_k=\operatorname{logit}(\lambda^{\star}_k)$. They provide mild regularization without ordering posterior transfer parameters.

\subsection{Sampling and Model Evaluation}
We monitor transfer parameters, profile locations and gaps, acquisition coefficients and gains, chapter effects, latent states, item difficulties, tree coefficients, profile indicators, and pointwise log-likelihood contributions. Block random-walk updates are used for the tree coefficients at each internal node and for the  transfer logits $(\eta_1,\ldots,\eta_K)$. Slice sampling is used for $\sigma_\epsilon$, $\sigma_{\beta0}$, $\sigma_G$, and $\sigma_\delta$.

WAIC is computed from the monitored pointwise log-likelihood terms
\[
\log p(Y_{ijt}\mid\theta_{it},b_{jt})
=Y_{ijt}\log p_{ijt}+(1-Y_{ijt})\log(1-p_{ijt}).
\]
Posterior predictive checks compare observed and replicated overall accuracy and the variance of chapter-level accuracy.

\section{Technical Properties of the Model}
\label{app:model_properties}

\subsection{Proof of Monotone Acquisition}

\begin{proof}[Proof of Proposition~\ref{prop:monotone_acquisition}]
The derivative of a B-spline expansion of degree $p$ is a B-spline expansion of degree $p-1$ whose coefficients are positive multiples of adjacent coefficient differences:
\[
f'_{km}(x)
=\sum_{\ell=1}^{L-1}c_\ell
(\beta_{km,\ell+1}-\beta_{km\ell})\widetilde B_\ell(x),
\]
where $\widetilde B_\ell(x)\geq 0$ and, with the knot vector indexed in the standard way, $c_\ell>0$. Nonnegative adjacent differences therefore imply $f'_{km}(x)\geq0$.
\end{proof}

\subsection{Proof of the Finite Curriculum-Path Representation}

\begin{proof}[Proof of Theorem~\ref{thm:path}]
Strict lower triangularity of $\mathbf W$ implies $\mathbf W^r=\mathbf 0$ for $r\geq T$, and the same holds for $\mathbf H_k=\lambda_k\mathbf W$. The vector recursion gives $(\mathbf I-\mathbf H_k)\boldsymbol\theta_i=\mathbf u_i+\boldsymbol\epsilon_i$. Nilpotency yields $(\mathbf I-\mathbf H_k)^{-1}=\sum_{r=0}^{T-1}\mathbf H_k^r$, establishing \eqref{eq:path_expansion}. Finally, $[\mathbf H_k^r]_{tc} =\lambda_k^r  [\mathbf W^r]_{tc}$ sums products of edge weights over directed paths of length $r$ from $c$ to $t$; summing over $r$ gives $\tau_{k,tc}$.
\end{proof}

The representation separates direct transfer, $\lambda_kW_{tc}$, from indirect propagation, $\sum_{r=2}^{T-1}\lambda_k^r[\mathbf W^r]_{tc}$. It also gives the model-based sensitivity
\[
\frac{\partial E(\theta_{it}\mid\mathbf x_i,z_i=k)}{\partial u_{ic}}=\tau_{k,tc},\qquad c<t.
\]
Because $\mathbf W$ is nonnegative and $\lambda_k \in (0,1)$,
\[
\tau_{k,tc}\geq0,
\qquad
\frac{\partial\tau_{k,tc}}{\partial\lambda_k}
=\sum_{r=1}^{T-1}r\lambda_k^{r-1}[\mathbf W^r]_{tc}\geq0.
\]
Thus, pairwise propagation and effective curriculum cohesion $C_k$ are non-decreasing in transfer persistence when the curriculum network is fixed.

\section{Simulation Study Details}
\label{app:simulation}

\subsection{Simulation Design}

Each simulated data set contained $N=198$ students observed across $T=9$
chapters, with chapter-specific item counts
\[
(J_1,\ldots,J_9)=(13,32,31,32,32,30,12,20,29),
\]
for a total of $45{,}738$ binary item responses. Chapters were assigned to
$M=4$ pedagogical topics using the mapping
\[
m(t)=(1,1,2,2,3,3,3,3,4).
\]
We repeated the data-generation and model-fitting procedure for $R=200$
independent replications.

The true population contained $K_{\mathrm{true}}=4$ learner profiles. Profile
membership was generated using the same balanced-tree stick-breaking
construction as in the proposed model. Five binary student characteristics were
generated for each student: gender, mathematics anxiety, memorization-oriented
beliefs, prior mathematics preparation, and prior coding experience. The
covariate distributions were chosen to resemble those in the motivating data.
The tree-branch coefficients were specified to create moderate, rather than
nearly deterministic, separation among the four profiles.

Student engagement was generated at the chapter level from three observed
components: submission frequency, average activity duration, and engagement
rate. Submission frequency was generated from a Poisson model, activity
duration from a Gamma distribution, and engagement rate from a Beta
distribution. These components were combined into a raw engagement measure,
\[
x^{\mathrm{raw}}_{it}
=
n_{it}d_{it}e_{it},
\]
and then transformed as
\[
x_{it}
=
\frac{\log(1+x^{\mathrm{raw}}_{it})-\bar{x}_{\log}}
{s_{\log}},
\]
where $\bar{x}_{\log}$ and $s_{\log}$ are the mean and standard deviation of
$\log(1+x^{\mathrm{raw}}_{it})$ over all students and chapters in the simulated
data set. This standardized engagement variable $x_{it}$ was used directly as
the spline input in the fitted model. Engagement depended on learner profile,
chapter, prior mathematics preparation, and mathematics anxiety, thereby
creating realistic dependence between engagement behavior and latent learning
patterns.

For learner profile $k$ and topic $m$, the true novel-acquisition function was
generated from a monotone concave exponential curve,
\[
f_{km}^{\mathrm{true}}(x)
=
a_k+c_m+
g_k^{\mathrm{profile}}g_m^{\mathrm{topic}}
\left[
1-\exp\left\{
-r_k^{\mathrm{profile}}r_m^{\mathrm{topic}}(x-x_{\min})
\right\}
\right],
\]
where $x_{\min}$ is the minimum standardized engagement value in the simulated
data set. The shift by $x_{\min}$ ensures that the exposure distance is
nonnegative over the observed support of standardized engagement. The profile
baseline values were
\[
(a_1,a_2,a_3,a_4)=(1.20,0.40,-0.30,-1.00),
\]
and the topic anchors were
\[
(c_1,c_2,c_3,c_4)=(0.10,0.05,-0.05,-0.10).
\]
Profiles differed in baseline acquisition, total gain, and acquisition rate,
whereas topics differed in their gain and rate multipliers. 

Curriculum-mediated transfer was generated using the curriculum similarity
matrix from the CourseKata application. For chapter $t$, the accumulated
contribution from earlier chapters was
\[
A_{it}
=
\sum_{c<t}\omega_{ct}\theta_{ic},
\]
where $\omega_{ct}$ is the normalized pedagogical similarity between earlier
chapter $c$ and current chapter $t$. The four learner profiles had transfer
parameters
\[
(\lambda_1,\lambda_2,\lambda_3,\lambda_4)
=
(0.60,0.55,0.50,0.45).
\]
The latent response propensity was generated as
\[
\theta_{it}
=
\lambda_{z_i}A_{it}
+
f_{z_i,m(t)}^{\mathrm{true}}(x_{it})
+
\delta_t
+
\varepsilon_{it},
\qquad
\varepsilon_{it}\sim N(0,0.65^2),
\]
where $\delta_t$ is a centered chapter effect. Chapter effects were generated
from a normal distribution with additional linear chapter-level structure and
then centered to satisfy
\[
\sum_{t=1}^{T}\delta_t=0.
\]
Finally, item difficulties were generated within chapter and centered so that
the chapter-specific item difficulties had mean zero. Binary item responses
followed the Rasch observation model
\[
Y_{ijt}\sim\operatorname{Bernoulli}(p_{ijt}),
\qquad
\operatorname{logit}(p_{ijt})=\theta_{it}-b_{jt}.
\]

For each simulated data set, we fitted \textsc{NAFI} with
\[
K\in\{2,4,8\}.
\]
The fitted acquisition functions were represented using $L=8$ cubic B-spline
basis functions. The spline was evaluated directly on standardized engagement
$x_{it}$, with boundary knots set to the observed minimum and maximum of
$x_{it}$ and inner knots placed at empirical quantiles. Monotonicity was
enforced through positive spline increments, and diminishing marginal gains
were encouraged through the normalized-increment construction described in
Section~\ref{sec:model}. The profile-level baselines were ordered to stabilize
profile labels.

For each fitted model and replication, we ran a single-chain NIMBLE MCMC sampler for 30,000 iterations, discarded the first 15,000 iterations as burn-in, and retained every tenth remaining draw. This yielded 1,500 posterior draws per fitted model.
Recovery was evaluated using profile classification accuracy, latent-state error, transfer-parameter recovery, chapter-effect recovery, covariate-effect recovery, learning-curve recovery over the central 90\% of the observed engagement support, WAIC, and posterior predictive checks.

\subsection{Evaluation Metrics}
For each replication, we evaluated recovery under the correctly specified
profile number $K=4$. Because mixture labels are arbitrary, profile-level
quantities were evaluated after aligning fitted profiles to the true profiles.
The alignment was chosen to maximize modal profile classification accuracy,
with the identity mapping preferred when tied. We report both raw and aligned
profile classification accuracy, but all profile-specific recovery summaries
below use the aligned labels.

We evaluated learner-profile recovery using overall aligned classification
accuracy, per-profile classification rates, posterior classification certainty,
and recovered modal profile sizes. Dynamic latent-state recovery was summarized
by the RMSE and MAE of posterior mean $\theta_{it}$ relative to the true
simulated values.

For scalar parameters, including transfer parameters $\lambda_k$, chapter
effects $\delta_t$, and balanced-tree stick-breaking covariate effects
$\gamma_{rp}$, we report Monte Carlo bias, RMSE, MAE, posterior standard
deviation, credible-interval coverage, and credible-interval width. For
engagement-acquisition curves, recovery was evaluated over the central 90\% of
the observed standardized engagement support. We report RMSE, MAE, bias,
pointwise coverage, and credible-interval width for each profile-topic
combination. We also separately examined curve recovery near the lower and
upper edges of the engagement support, where spline estimates are typically
less stable.

Posterior predictive checks were based on two summaries: overall response
accuracy and the variance of chapter-level accuracy. Posterior predictive
$p$-values near 0.5 indicate that the fitted model reproduces these empirical
features without systematic over- or under-dispersion.

\subsection{Simulation results}
\subsubsection{Selection of the number of learner profiles}
We first examined whether the simulation results supported the true number of learner profiles. 
The data-generating model contained $K_{\mathrm{true}}=4$ profiles. 
For each simulated data set, we compared models with $K\in\{2,4,8\}$ and further considered a restricted no-transfer specification under $K=4$ with $\lambda=0$. 
The purpose of these comparisons was to evaluate whether the fitted model could distinguish under-extraction of learner heterogeneity, unnecessary over-partitioning, and misspecification of curriculum-mediated transfer.

Table~\ref{tab:sim_model_selection} summarizes the Monte Carlo means of the main model comparison and diagnostic quantities across the simulation replications. 
The correctly specified $K=4$ model achieved the smallest mean WAIC among the profile-number specifications. 
The $K=2$ model had a substantially larger mean WAIC, with an increase of 52.88 relative to $K=4$, indicating that two profiles were insufficient to capture the heterogeneity generated by the four-profile data-generating process. 
Although the $K=2$ model had a high mean maximum posterior profile probability, this mainly reflects the fact that it partitions students into fewer, broader classes; it should therefore not be interpreted as better recovery of the true latent structure.

The $K=8$ model produced a mean WAIC nearly identical to that of the $K=4$ model, but slightly larger by 0.60 on average. Thus, doubling the number of profiles did not yield a meaningful predictive improvement. Given the increased interpretive complexity of the eight-profile solution,
the simulation supports the more parsimonious $K=4$ specification.
Thus, the model comparison indicates that the proposed approach avoids the clear underfitting associated with $K=2$ while not requiring unnecessary over-partitioning beyond the true four-profile structure.

The restricted transfer specifications provide additional information about the dynamic component of the model. 
Removing curriculum-mediated transfer by setting $\lambda=0$ led to a noticeably worse mean WAIC, increasing by 65.37 relative to the selected $K=4$ model. 
It also degraded recovery of the latent states, chapter effects, and engagement-acquisition curves. 
The large increase in curve error under the no-transfer model suggests that, when curriculum-mediated dependence is omitted, the acquisition functions and chapter effects compensate for the missing dynamic structure, leading to distorted estimates of engagement-associated acquisition.
This confirms that curriculum-mediated transfer is an important component of the data-generating mechanism and is needed for accurate recovery.


Overall, the simulation model-selection results support the use of $K=4$ as the main specification. 
The $K=4$ model balances predictive fit, correct recovery of the true latent-profile structure, and interpretability. 
The subsequent simulation results therefore focus on recovery under the correctly specified $K=4$ model.

\begin{table}[H]
\centering
\small
\begin{threeparttable}
\caption{Simulation model comparison and restricted transfer checks.}
\label{tab:sim_model_selection}
\begin{tabularx}{\textwidth}{
>{\raggedright\arraybackslash}X
>{\raggedright\arraybackslash}X
rrrrr}
\toprule
Specification 
& WAIC
& Max prob.
& Min. size
& $\theta$ RMSE
& Ch. RMSE
& Curve RMSE \\
\midrule

$K=4$
& 35256.79
& 0.954
& 35.43
& 0.496
& 0.119
& 0.127 \\

$K=2$
& 35309.67
& 0.963
& 84.57
& 0.519
& 0.234
& -- \\

$K=8$
& 35257.39
& 0.594
& 2.805
& 0.499
& 0.069
& -- \\

$K=4$, $\lambda=0$
& 35322.16
& 0.983
& 36.67
& 0.527
& 0.323
& 0.937 \\

\bottomrule
\end{tabularx}

\begin{tablenotes}[flushleft]
\footnotesize
\item Note. Entries: Monte Carlo means across simulation replications. 
Max prob.: the average maximum posterior profile probability. 
Min. size denotes the mean of the smallest modal profile size across replications. 
Ch. RMSE: the root mean square error between true values and posterior mean estimates of $\delta_t$ at the chapter level.
Curve RMSE: reported for specifications where profile-specific curve recovery was computed. 
\end{tablenotes}
\end{threeparttable}
\end{table}

These results justify focusing the remaining simulation analysis on recovery under the correctly specified $K=4$ model, while also showing that the proposed framework detects underfitting, avoids a clear need for over-partitioning, and benefits from including curriculum-mediated transfer.

We next assessed overall fit and posterior predictive calibration for the correctly specified $K=4$ model. 
Table~\ref{tab:sim_fit_summary} summarizes WAIC, effective complexity, posterior predictive checks, and runtime across the 200 replications. 
The posterior predictive checks used the same discrepancy statistics as in the empirical analysis: overall response accuracy and the variance of chapter-level accuracy.

\begin{figure}[t]
    \centering
    \includegraphics[width=0.75\textwidth]{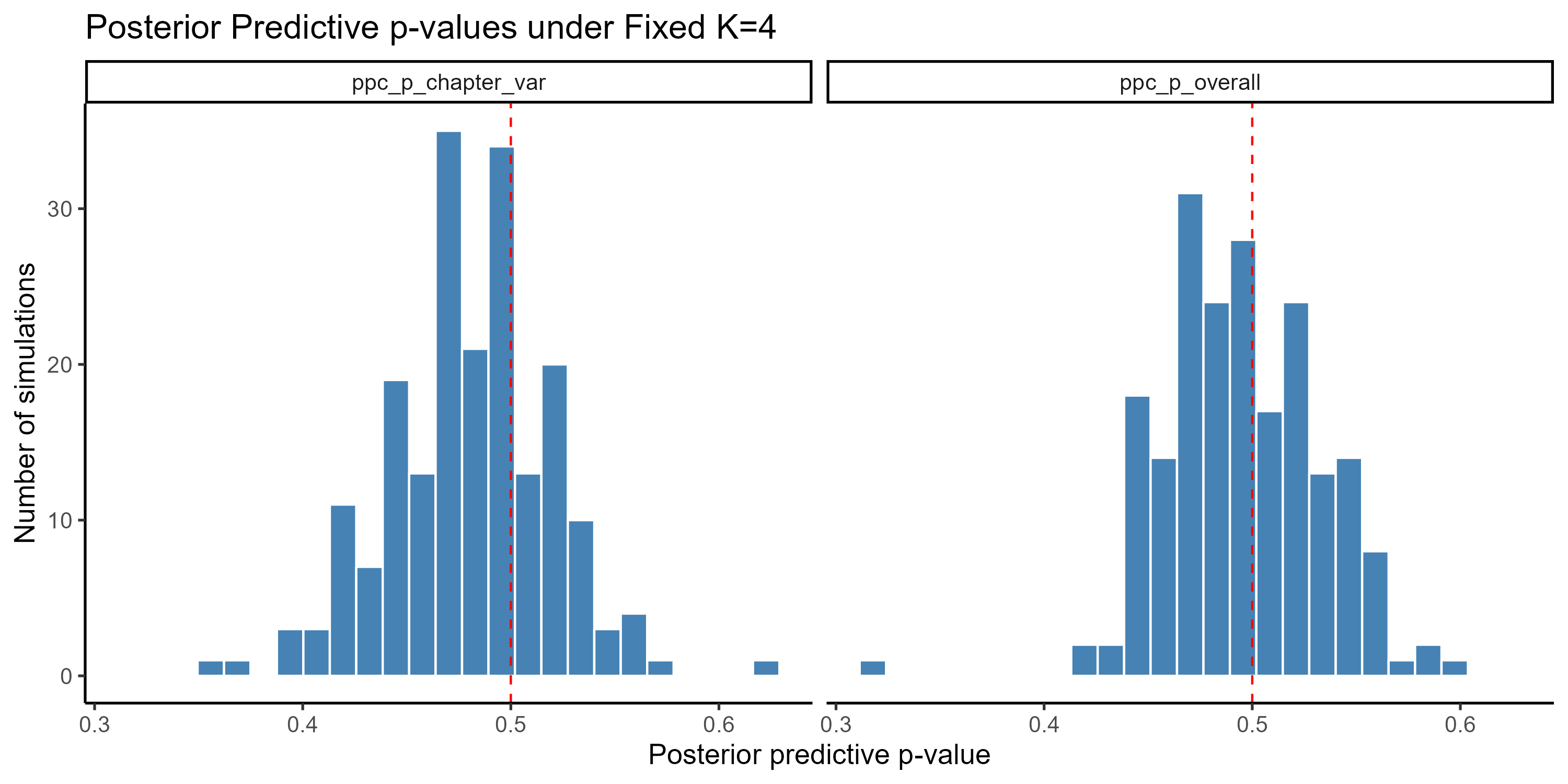}
    \caption{Posterior predictive $p$-value distributions under $K=4$.}
    \label{fig:sim_ppc_pvalues}
\end{figure}

The posterior predictive $p$-values were centered near 0.5 for both summaries, with mean values of 0.495 for overall accuracy and 0.480 for chapter-level accuracy variation. 
This indicates that, when the data were generated from the proposed dynamic profile model, the fitted model reproduced both the overall response rate and the between-chapter variation in accuracy without systematic misfit. 
The distributions of posterior predictive $p$-values are shown in Figure~\ref{fig:sim_ppc_pvalues}.

\begin{table}[t]
\centering
\begin{threeparttable}
\caption{Simulation fit and posterior predictive summaries under $K=4$.}
\label{tab:sim_fit_summary}
\begin{tabular}{lrrrrr}
\toprule
Metric & Mean & SD & Median & 2.5\% & 97.5\% \\
\midrule
WAIC 
& 35256.79 
& 1693.63 
& 35331.66 
& 31721.93 
& 38025.75 \\

elpd WAIC 
& -17628.39 
& 846.81 
& -17665.83 
& -19012.87 
& -15860.96 \\

$p_{\mathrm{WAIC}}$ 
& 1071.64 
& 38.50 
& 1074.20 
& 997.16 
& 1141.02 \\

PPC $p$-value: overall accuracy 
& 0.495 
& 0.038 
& 0.490 
& 0.440 
& 0.565 \\

PPC $p$-value: chapter variance 
& 0.480 
& 0.039 
& 0.480 
& 0.405 
& 0.555 \\

Runtime minutes 
& 85.88 
& 3.20 
& 85.56 
& 81.08 
& 91.75 \\
\bottomrule
\end{tabular}
\begin{tablenotes}[flushleft]
\footnotesize
\item Note. Summaries are computed over 200 Monte Carlo replications under the correctly specified $K=4$ model. 
Posterior predictive checks use overall response accuracy and the variance of chapter-level accuracy as discrepancy statistics. 
Posterior predictive $p$-values near 0.5 indicate adequate calibration with respect to the corresponding summary.
\end{tablenotes}
\end{threeparttable}
\end{table}

\subsubsection{Recovery of learner profile}
We next evaluated whether the model recovered the true learner profiles. Because
profile labels are arbitrary in mixture models, we aligned fitted profile labels
to the true labels within each replication before computing profile-specific
recovery metrics. Table~\ref{tab:sim_profile_recovery} summarizes the raw and
aligned classification accuracy under the correctly specified four-profile
solution.

Before label alignment, the mean modal classification accuracy was relatively
low, $\bar{A}_{\mathrm{raw}} = 0.825$, with substantial variability across
replications. After label alignment, the mean classification accuracy increased
to $\bar{A}_{\mathrm{aligned}} = 0.955$, with a much smaller standard deviation.
The median alignment gain was zero, indicating that many replications already
had stable label ordering. However, the mean alignment gain was $0.130$, and
$26\%$ of replications showed gains larger than $0.20$, suggesting that
non-negligible label switching occurred in a subset of replications. Thus,
label alignment was important for obtaining an interpretable assessment of
profile recovery.

\begin{table}[t]
\centering
\begin{threeparttable}
\caption{Learner-profile recovery under $K=4$.}
\label{tab:sim_profile_recovery}
\begin{tabular}{lrrr}
\toprule
Metric & Mean & SD & MCSE \\
\midrule
Raw classification accuracy
  & 0.825 & 0.236 & 0.017  \\
Aligned classification accuracy
  & 0.955 & 0.027 & 0.002  \\
\bottomrule
\end{tabular}
\begin{tablenotes}[flushleft]
\footnotesize
\item Note. Profile labels were aligned within each replication by maximizing
modal classification accuracy. MCSE denotes Monte Carlo standard error across
replications.
\end{tablenotes}
\end{threeparttable}
\end{table}

\begin{figure}[H]
    \centering
    \includegraphics[width=0.55\textwidth]{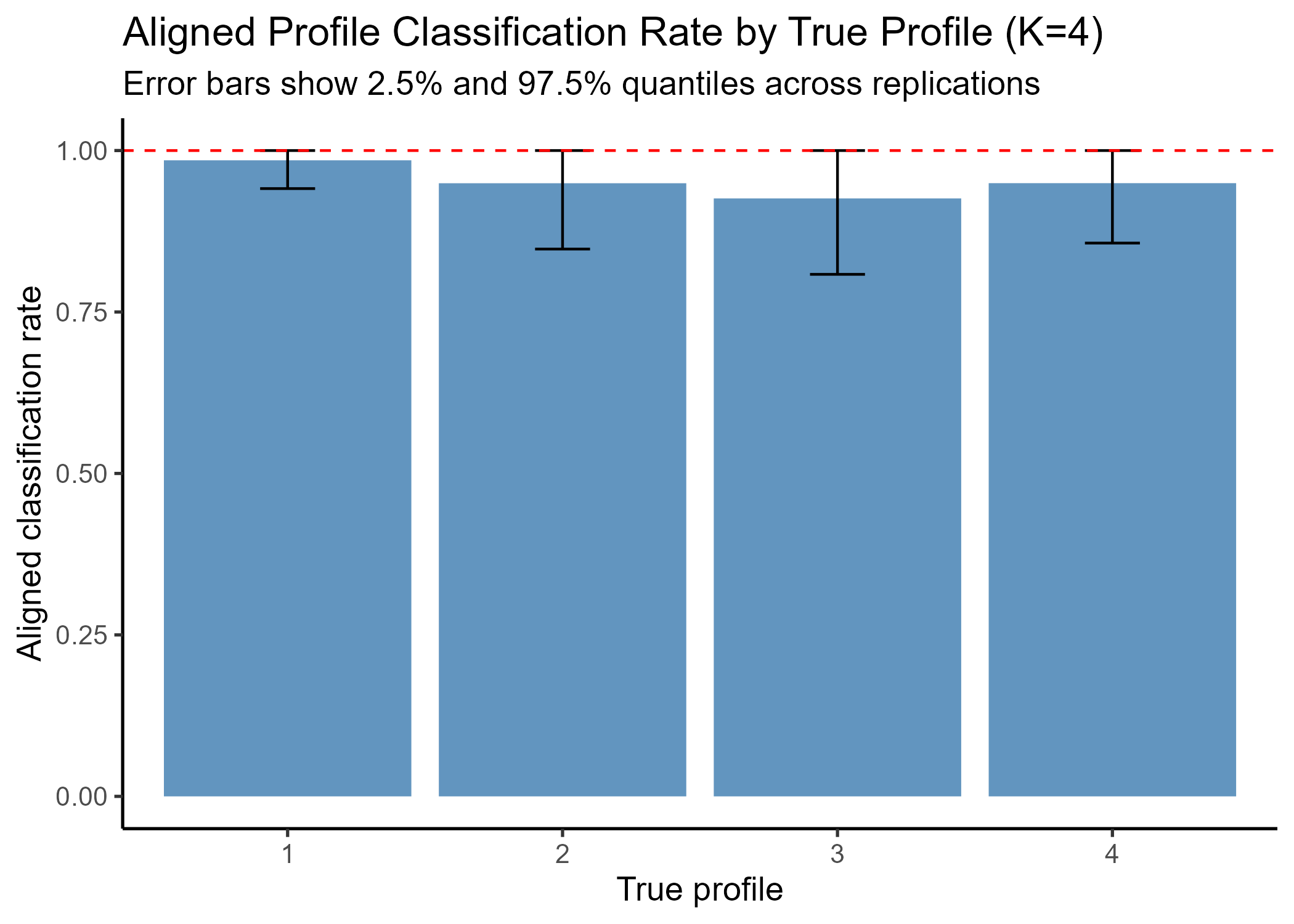}
    \caption{Aligned profile classification rates by true profile under $K=4$.
    Error bars show the 2.5\% and 97.5\% empirical quantiles across the 200
    simulation replications.}
    \label{fig:sim_profile_rate}
\end{figure}

Figure~\ref{fig:sim_profile_rate} reports aligned classification rates by true
profile. Classification performance was highest for Profile~1, whose mean
aligned classification rate was $0.985$. Profiles~2 and~4 also showed strong
recovery, with mean aligned classification rates of $0.949$ and $0.949$,
respectively. Profile~3 was the most difficult to recover, with a mean
classification rate of $0.926$ and a wider empirical interval across
replications. This pattern suggests that Profile~3 had greater overlap with
neighboring profiles in the latent acquisition and engagement processes.
Nevertheless, all four profiles were recovered with high average accuracy, and
the lower bounds of the empirical $95\%$ intervals remained above approximately
$0.81$ for all profiles.

Table~\ref{tab:sim_profile_rate_by_true} provides the profile-specific recovery
results. The mean maximum posterior profile probabilities closely tracked the
classification rates, ranging from $0.930$ for Profile~3 to $0.980$ for
Profile~1. This indicates that the model not only assigned most learners to the
correct profile, but also did so with high posterior certainty.

\begin{table}[t]
\centering
\begin{threeparttable}
\caption{Aligned classification rates by true learner profile under $K=4$.}
\label{tab:sim_profile_rate_by_true}
\begin{tabular}{lrrrrrrr}
\toprule
True profile & Mean $n$ & Mean rate & SD & MCSE & 2.5\% & 97.5\% & Mean max prob. \\
\midrule
1 & 59.28 & 0.985 & 0.027 & 0.002 & 0.941 & 1.000 & 0.980 \\
2 & 54.64 & 0.949 & 0.039 & 0.003 & 0.847 & 1.000 & 0.952 \\
3 & 47.69 & 0.926 & 0.082 & 0.006 & 0.808 & 1.000 & 0.930 \\
4 & 36.39 & 0.949 & 0.083 & 0.006 & 0.857 & 1.000 & 0.945 \\
\bottomrule
\end{tabular}
\begin{tablenotes}[flushleft]
\footnotesize
\item Note. Rates are computed after aligning fitted profile labels to the true
labels within each replication. Mean max prob. denotes the average maximum
posterior probability for students whose true profile is the corresponding row.
\end{tablenotes}
\end{threeparttable}
\end{table}

The recovered modal profile sizes were also close to the true profile sizes.
As shown in Table~\ref{tab:sim_profile_size}, the mean recovered modal sizes
were $58.40$, $53.70$, $47.85$, and $38.06$ for Profiles~1--4, respectively,
compared with true mean sizes of $59.28$, $54.64$, $47.69$, and $36.39$.
The corresponding mean differences were small, ranging from $-0.94$ to $1.67$
students. Therefore, when $K$ was correctly specified, the fitted model did not
systematically collapse or split the four true learner groups.

\begin{table}[t]
\centering
\begin{threeparttable}
\caption{True and recovered modal profile sizes under $K=4$.}
\label{tab:sim_profile_size}
\begin{tabular}{lrrrr}
\toprule
Profile & True mean $n$ & True SD & Modal mean $n$ & Modal SD \\
\midrule
1 & 59.28 & 6.66 & 58.40 & 7.32  \\
2 & 54.64 & 6.21 & 53.70 & 8.72  \\
3 & 47.69 & 6.54 & 47.85 & 8.08  \\
4 & 36.39 & 5.70 & 38.06 & 8.50\\
\bottomrule
\end{tabular}
\begin{tablenotes}[flushleft]
\footnotesize
\item Note. True size refers to the number of students generated from each true
profile. Modal size refers to the number of students assigned to each aligned
profile by modal posterior classification.
\end{tablenotes}
\end{threeparttable}
\end{table}

\begin{figure}[H]
    \centering
    \includegraphics[width=0.55\textwidth]{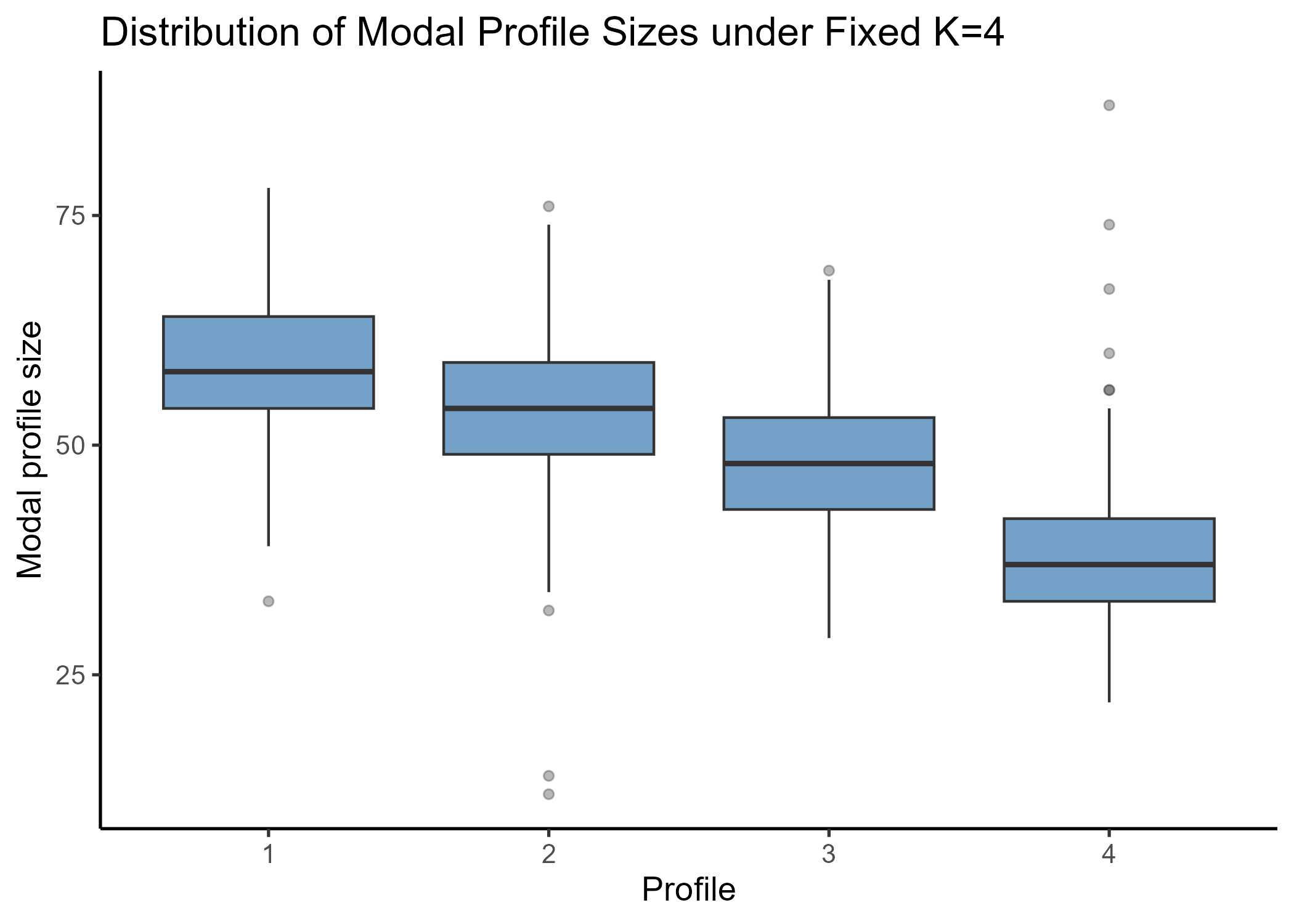}
    \caption{Distribution of recovered modal profile sizes under $K=4$.}
    \label{fig:sim_profile_size}
\end{figure}

Figure~\ref{fig:sim_profile_size} further displays the distribution of recovered
modal profile sizes. The central tendency of each recovered profile size closely
matched the corresponding true profile size. Some outlying replications were
observed, especially for Profiles~2 and~4, but these did not indicate systematic
profile collapse. Instead, they reflect occasional sampling and classification
variability across replications.

\begin{figure}[t]
    \centering
    \includegraphics[width=0.55\textwidth]{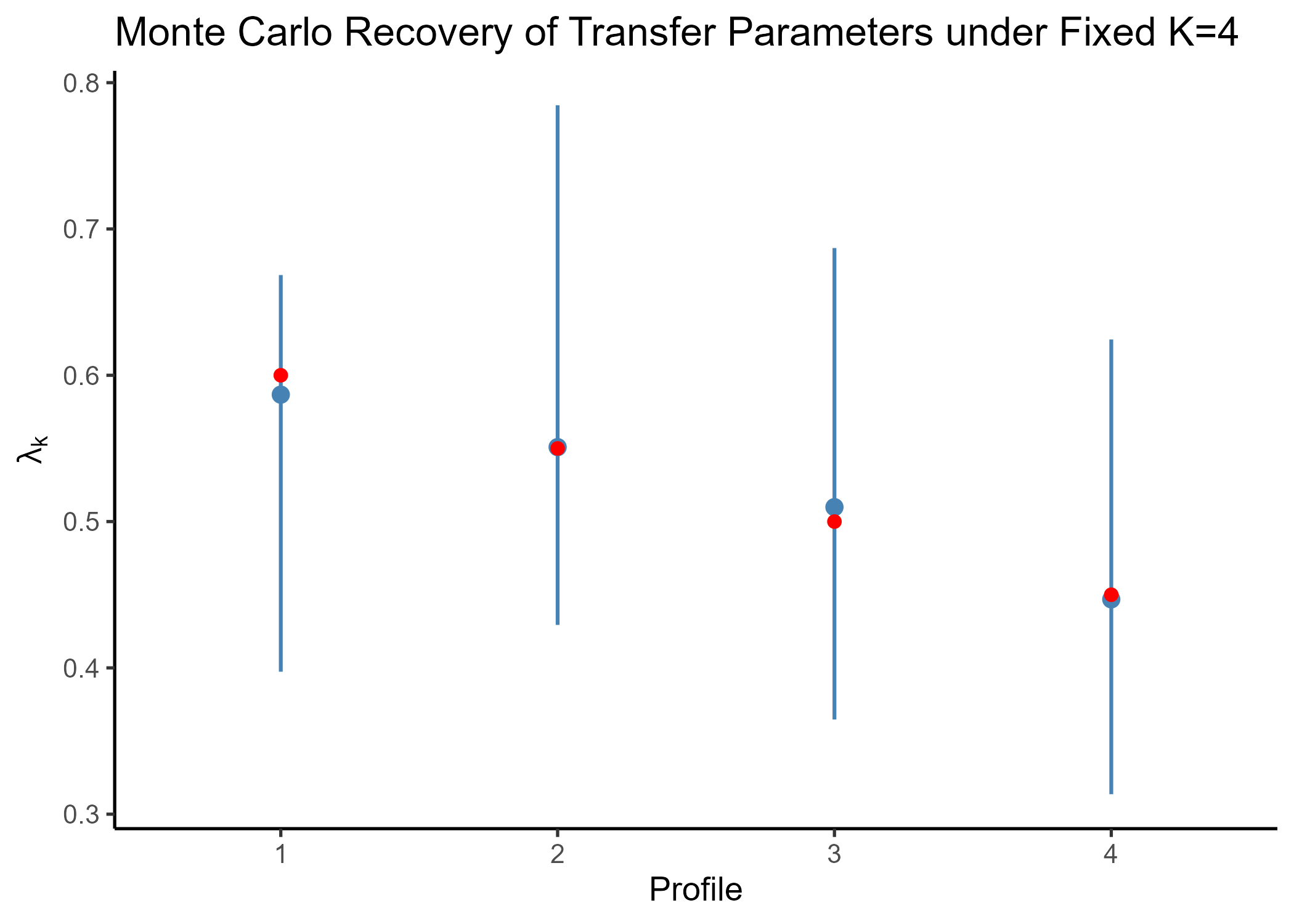}
    \caption{Monte Carlo recovery of profile-specific transfer parameters under $K=4$. Red points denote true values; blue points denote average posterior mean estimates across replications; blue intervals show Monte Carlo variation in posterior mean estimates across replications.}
    \label{fig:sim_lambda_recovery}
\end{figure}

Table~\ref{tab:sim_parameter_recovery} summarizes recovery of the main dynamic
and structural parameters under $K=4$. The profile-specific transfer
parameters $\lambda_k$ were recovered with negligible average signed bias
($-0.001$) and small average RMSE ($0.080$). The empirical coverage of the
nominal 95\% posterior credible intervals was 0.934 on average, slightly below
but close to the nominal level. As shown in
Figure~\ref{fig:sim_lambda_recovery}, the posterior mean estimates closely
tracked the true transfer parameters across the four profiles. These results
indicate that the model can identify curriculum-mediated transfer when the
curriculum similarity matrix is known and the number of profiles is correctly
specified.

\begin{figure}[t]
    \centering
    \includegraphics[width=0.55\textwidth]{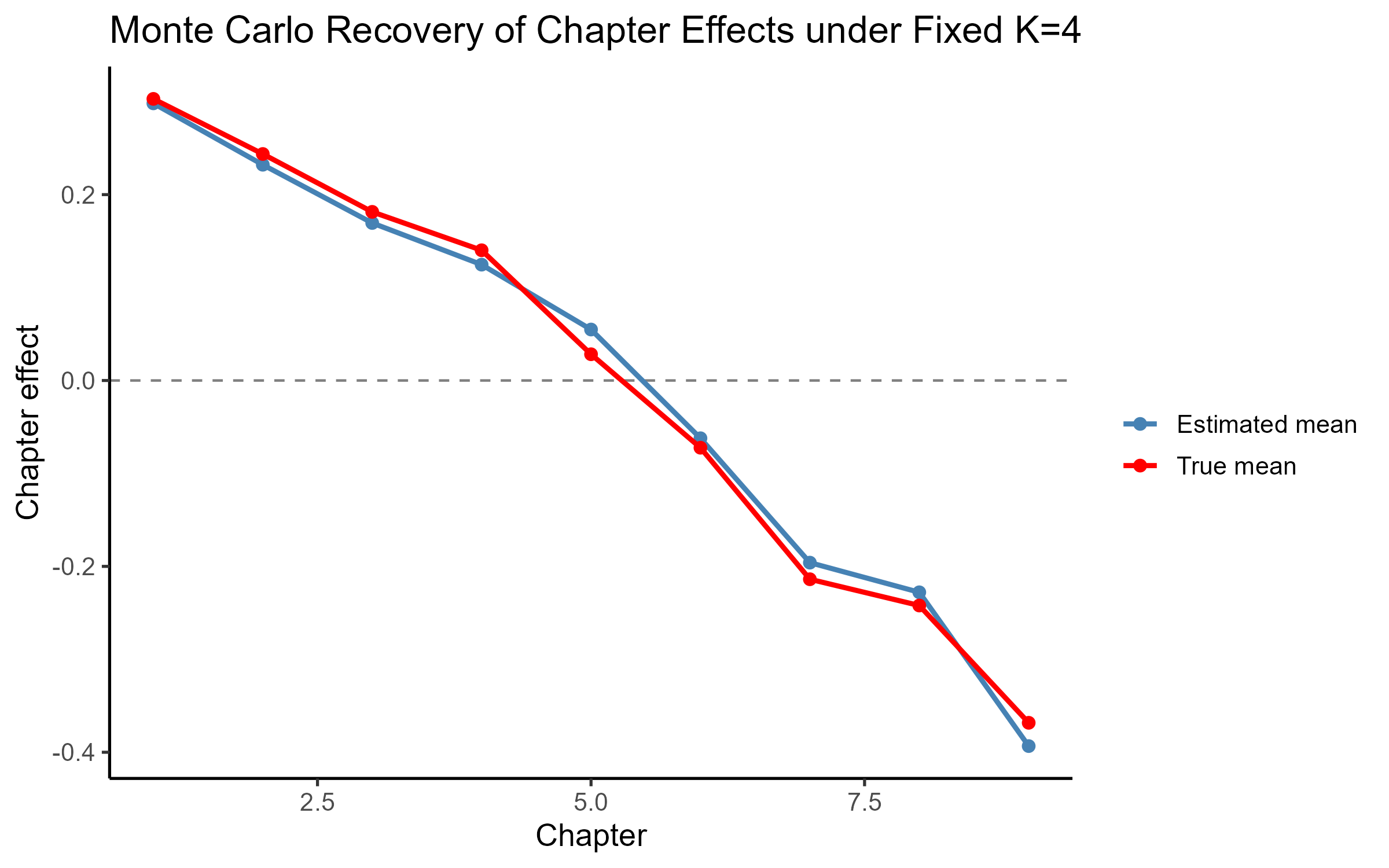}
    \caption{Monte Carlo recovery of centered chapter effects under $K=4$. Red points denote true chapter effects; blue points denote average posterior mean estimates across replications.}
    \label{fig:sim_chapter_recovery}
\end{figure}

The centered chapter effects were also well recovered. The average signed bias
was essentially zero, and the average RMSE and MAE were 0.119 and 0.082,
respectively. Coverage was mildly conservative on average, with empirical
coverage of 0.975 and mean credible-interval width of 0.463. As shown in
Figure~\ref{fig:sim_chapter_recovery}, the estimated chapter-effect trajectory
closely tracked the true trajectory across chapters, including the transition
from positive early-chapter effects to negative later-chapter effects. This
suggests that the model can distinguish residual chapter-level variation from
engagement-related acquisition and curriculum-mediated transfer.

\begin{figure}[t]
    \centering
    \includegraphics[width=0.8\textwidth]{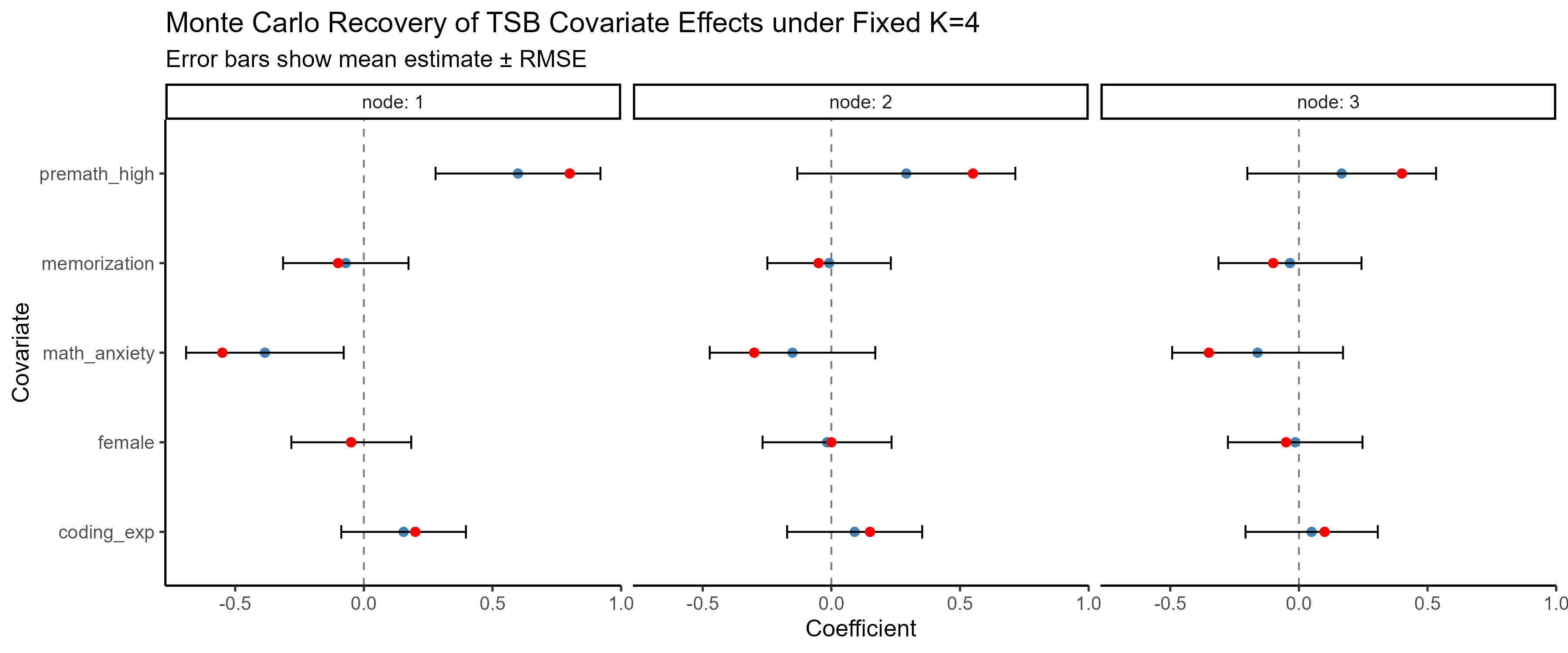}
    \caption{Monte Carlo recovery of balanced-tree stick-breaking covariate effects under $K=4$. Red points denote true values; blue points denote average posterior mean estimates across replications. Error bars show posterior mean estimates plus or minus one RMSE.}
    \label{fig:sim_gamma_recovery}
\end{figure}

The balanced-tree stick-breaking covariate effects were estimated with greater
uncertainty than the lower-dimensional dynamic parameters. Their average RMSE
was 0.295 and the mean credible-interval width was 1.266, reflecting the fact
that these covariates affect the likelihood only indirectly through latent
profile membership. Nevertheless, the average signed bias was small
($-0.006$), and empirical coverage was close to nominal on average
(0.960). The main directional patterns were recovered, especially for covariates
with larger true effects, such as prior mathematics preparation and mathematics
anxiety. At the same time, the estimates showed some shrinkage toward zero for
the largest effects, particularly for the prior mathematics preparation effects
and the negative mathematics-anxiety effects.
Monte Carlo recovery of the balanced-tree stick-breaking covariate effects is shown in Figure~\ref{fig:sim_gamma_recovery}.

The total gain parameters $G_{km}$ were recovered with moderate accuracy. The
average signed bias was $-0.039$, with average RMSE 0.118 and MAE 0.090.
Coverage was high on average (0.979), although the largest gain parameters in
the first profile showed some downward bias. This pattern indicates that the
model recovered the overall profile-topic gain structure well, while tending to
slightly attenuate the largest gain effects.

Finally, posterior mean latent states $\theta_{it}$ showed moderate prediction
error across students and chapters, with average RMSE 0.496 and average MAE
0.384. Given that the latent states are inferred from binary item-response data,
this level of error indicates that the model recovered individual-level latent
response propensities with reasonable accuracy.

\begin{table}[t]
\centering
\begin{threeparttable}
\caption{Monte Carlo recovery of key parameters under $K=4$.}
\label{tab:sim_parameter_recovery}
\begin{tabular}{lrrrrr}
\toprule
Parameter class & Bias & RMSE & MAE & Coverage & Mean CI width \\
\midrule
Transfer parameters $\lambda_k$ & -0.001 & 0.080 & 0.059 & 0.934 & 0.283 \\
Chapter effects $\delta_t$ & 0.000 & 0.119 & 0.082 & 0.975 & 0.463 \\
TSB covariate effects $\gamma$ & -0.006 & 0.295 & 0.232 & 0.960 & 1.266 \\
Total gain parameters $G_{km}$ & -0.039 & 0.118 & 0.090 & 0.979 & 0.581 \\
Latent states $\theta_{it}$ & -- & 0.496 & 0.384 & -- & -- \\
\bottomrule
\end{tabular}
\begin{tablenotes}[flushleft]
\footnotesize
\item Note. Entries summarize Monte Carlo performance across replications and,
where applicable, across profiles, chapters, topics, nodes, or covariates.
Bias is the signed average bias. Coverage is the empirical coverage of nominal
95\% posterior credible intervals. For $\theta_{it}$, the table reports RMSE
and MAE of posterior means rather than credible-interval coverage.
\end{tablenotes}
\end{threeparttable}
\end{table}

\subsubsection{Recovery of engagement-acquisition curves}

We then evaluated recovery of the nonlinear engagement-acquisition functions.
The data were generated from monotone
concave exponential curves, whereas the fitted model used monotone B-splines.
Thus, recovery assesses whether the spline representation can approximate the
main acquisition patterns without being given the true functional form.

\begin{figure}[H]
    \centering
    \includegraphics[width=0.8\textwidth]{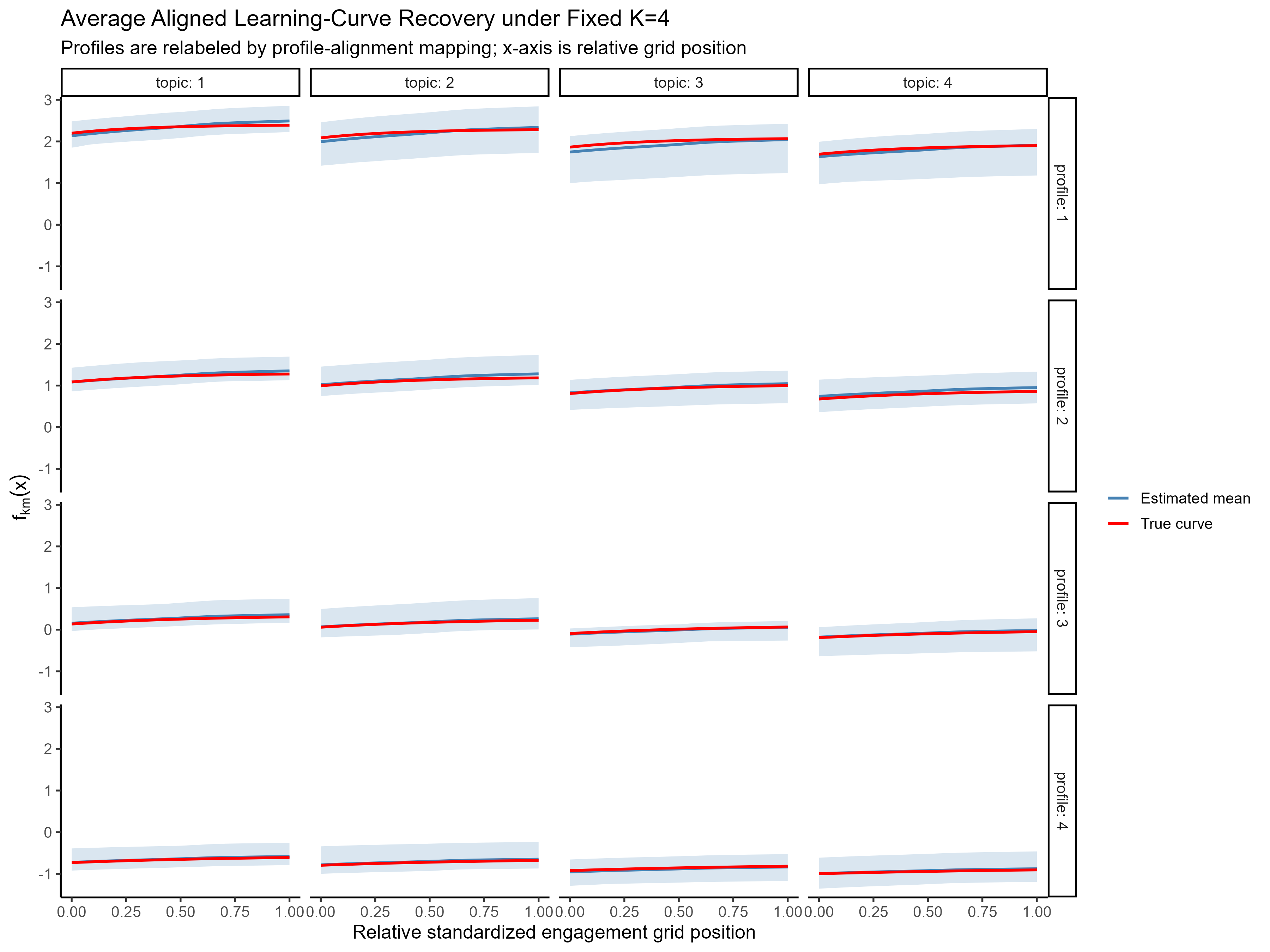}
    \caption{Average aligned recovery of engagement-acquisition curves under $K=4$. Red curves denote the true acquisition functions, blue curves denote Monte Carlo average posterior mean estimates, and shaded bands show Monte Carlo variation across replications.}
    \label{fig:sim_average_curves}
\end{figure}

Figure~\ref{fig:sim_average_curves} compares the average aligned estimated
curves with the true curves for each profile-topic combination. The estimated
curves closely followed the true curves over the central region of the observed
engagement support. The model recovered both the ordering of profiles and the
topic-specific differences in acquisition gains. Recovery was particularly
strong for Profiles~2--4, whereas Profile~1 showed somewhat larger errors,
especially for topics with larger acquisition gains and stronger curvature.

Table~\ref{tab:sim_curve_recovery} summarizes aligned curve recovery by learner
profile, averaging across the four topics. Overall, the mean RMSE was 0.127 and
the mean MAE was 0.124 across the 16 profile-topic curves. The average signed
bias was small, 0.008, indicating little systematic over- or under-estimation
after alignment. Pointwise credible-interval coverage was generally close to
the nominal level, with an overall mean coverage of 0.964. Coverage was slightly
lower for Profile~1, where the acquisition curves had larger gains and were
therefore more difficult to approximate, but remained close to 0.95 on average.

The uncertainty calibration results in Table~\ref{tab:sim_curve_calibration}
show a similar pattern. The mean pointwise coverage was close to nominal across
profiles. The uncertainty ratio, defined as the empirical standard deviation of
posterior mean estimates divided by the average posterior standard deviation,
was close to one for Profiles~1 and~2 but larger for Profiles~3 and~4. This
suggests that uncertainty for lower-performing profiles may be somewhat
underestimated relative to Monte Carlo variability, although the resulting
credible intervals still achieved high pointwise coverage. Overall, the
simulation indicates that the monotone spline representation can recover the
main engagement-acquisition patterns under functional misspecification.

\begin{table}[t]
\centering
\begin{threeparttable}
\caption{Aligned engagement-acquisition curve recovery under $K=4$.}
\label{tab:sim_curve_recovery}
\begin{tabular}{lrrrrrr}
\toprule
Profile & RMSE & MAE & Bias & Coverage & Mean CI width & Max abs. error \\
\midrule
1 & 0.166 & 0.158 & -0.032 & 0.945 & 0.827 & 0.212 \\
2 & 0.124 & 0.120 & 0.044 & 0.967 & 0.667 & 0.151 \\
3 & 0.101 & 0.099 & 0.014 & 0.972 & 0.550 & 0.114 \\
4 & 0.118 & 0.117 & 0.008 & 0.973 & 0.628 & 0.128 \\
\midrule
Overall & 0.127 & 0.124 & 0.008 & 0.964 & 0.668 & 0.152 \\
\bottomrule
\end{tabular}
\begin{tablenotes}[flushleft]
\footnotesize
\item Note. Curve recovery is computed after aligning fitted profiles to true
profiles within each replication. Entries average recovery metrics across
topics within each profile. The overall row averages across all 16
profile-topic curves. Coverage is pointwise coverage of nominal 95\% posterior
credible intervals over the central 90\% of the observed standardized
engagement support.
\end{tablenotes}
\end{threeparttable}
\end{table}

\begin{table}[t]
\centering
\begin{threeparttable}
\caption{Pointwise calibration of aligned engagement-acquisition curves under $K=4$.}
\label{tab:sim_curve_calibration}
\begin{tabular}{lrrrr}
\toprule
Profile & Pointwise RMSE & Pointwise coverage & Mean uncertainty ratio & Median uncertainty ratio \\
\midrule
1 & 0.239 & 0.945 & 1.088 & 1.090 \\
2 & 0.168 & 0.967 & 0.945 & 0.947 \\
3 & 0.184 & 0.972 & 1.327 & 1.331 \\
4 & 0.238 & 0.973 & 1.506 & 1.509 \\
\midrule
Overall & 0.207 & 0.964 & 1.217 & 1.219 \\
\bottomrule
\end{tabular}
\begin{tablenotes}[flushleft]
\footnotesize
\item Note. Pointwise RMSE and coverage are averaged over engagement-grid
points, topics, and replications within each profile. The uncertainty ratio is
the empirical standard deviation of posterior mean estimates divided by the
average posterior standard deviation approximated from the 95\% credible
interval width. Values near one indicate close agreement between Monte Carlo
variation and posterior uncertainty.
\end{tablenotes}
\end{threeparttable}
\end{table}

\subsubsection{Summary of simulation findings}

The simulation study supports the use of \textsc{NAFI} for recovering the main
learning mechanisms targeted by the proposed model. The model comparison results
favored the correctly specified $K=4$ solution over the under-extracted $K=2$
model, while the over-extracted $K=8$ model provided no meaningful predictive
gain. The no-transfer model performed substantially worse, confirming that
curriculum-mediated transfer is needed to recover the simulated learning
process.

Under $K=4$, the model reproduced the main posterior predictive features
of the data, recovered the four learner profiles with high aligned
classification accuracy, and produced modal profile sizes close to the true
profile sizes. Transfer parameters, chapter effects, total gain parameters, and
latent response propensities were recovered with reasonable accuracy. The
balanced-tree stick-breaking covariate effects were estimated with greater
uncertainty, which is expected because these parameters affect the response data
indirectly through latent profile membership.

The nonlinear engagement-acquisition curves were also recovered well. The fitted curves preserved the main
profile-topic ordering, showed modest average error, and achieved close-to-
nominal pointwise coverage over the observed engagement support. The main
limitations were occasional label switching, weaker separation for one profile,
and greater uncertainty for covariate effects and some lower-performing
profile-specific curves.

Overall, the simulation results indicate that \textsc{NAFI} can recover latent
learner heterogeneity, curriculum-mediated transfer, and nonlinear acquisition
patterns in a CourseKata-like setting. These findings support the empirical
interpretation of learner profiles and acquisition curves, while motivating a
cautious interpretation of fine-grained transfer differences and covariate
effects.
\section{CourseKata Model Selection and Structural Checks}
\label{app:coursekata_model_selection}

Table~\ref{tab:model_comparison} reports the comparison of models with $K=2$, $K=4$, and $K=8$ latent profiles.
The $K=8$ model achieved the smallest WAIC, but this apparent improvement in predictive fit came at the cost of poor profile stability: the model contained an empty profile and had a low mean maximum posterior profile probability.
This suggests that the additional profiles in the $K=8$ solution may reflect over-partitioning rather than substantively interpretable learner heterogeneity.

By contrast, the $K=2$ model was stable and well separated, but its WAIC was substantially larger, indicating that two profiles were insufficient to capture the observed heterogeneity in learner performance, engagement, and background characteristics.
The $K=4$ model provided the best balance between predictive fit, classification quality, and substantive interpretability.
Its WAIC was within 6 points of the minimum, it had no empty profiles, and its mean maximum posterior profile probability was 0.800.
We therefore use the $K=4$ model as the primary specification for substantive interpretation.

\begin{table}[t]
\centering
\begin{threeparttable}
\caption{CourseKata model comparison across profile specifications.}
\label{tab:model_comparison}
\begin{tabular}{rrrrrrc}
\toprule
$K$
& WAIC
& $\overline{\max_k p(z_i=k\mid \text{data})}$
& Min. size
& Empty
& Max PSRF
& Min ESS \\
\midrule
2
& 38638.71
& 0.894
& 52
& 0
& 1.042
& 160.9  \\
4
& 38610.16
& 0.800
& 12
& 0
& 1.012
& 302.8  \\
8
& 38604.20
& 0.422
& 0
& 1
& 1.087
& 154.0  \\
\bottomrule
\end{tabular}
\begin{tablenotes}[flushleft]
\footnotesize
\item Note. WAIC is the widely applicable information criterion, with smaller values indicating better expected predictive fit.
The posterior classification certainty is summarized by the average maximum posterior profile probability.
\end{tablenotes}
\end{threeparttable}
\end{table}

We conducted two ablation checks to evaluate the roles of learner-profile heterogeneity and curriculum-mediated transfer.
As shown in Table~\ref{tab:ablation_checks}, the no-profile model $(K=1)$ had substantially worse WAIC than the selected $K=4$ model, supporting the need to model latent learner heterogeneity.
The no-transfer model $(\lambda=0)$ also had worse WAIC, indicating that curriculum-mediated dependence contributes to predictive fit.
Together, these results support the two central structural components of the empirical model: learner profiles and curriculum-based transfer.

\begin{table}[t]
\centering
\begin{threeparttable}
\caption{CourseKata ablation checks for learner heterogeneity and curriculum-mediated transfer.}
\label{tab:ablation_checks}
\begin{tabular}{lrrrrr}
\toprule
Model specification
& $K$
& WAIC
& $\Delta$WAIC
& PPC overall
& PPC chapter var. \\
\midrule
Selected model
& 4
& 38610.16
& 0.00
& 0.455
& 0.420 \\
No transfer: $\lambda=0$
& 4
& 38618.77
& 8.61
& 0.455
& 0.485 \\
No learner profiles
& 1
& 38675.74
& 65.58
& 0.525
& 0.480 \\
\bottomrule
\end{tabular}
\begin{tablenotes}[flushleft]
\footnotesize
\item Note. $\Delta$WAIC is computed relative to the selected $K=4$ model.
The no-transfer model removes curriculum-mediated dependence, while the $K=1$ model removes latent learner-profile heterogeneity.
\end{tablenotes}
\end{threeparttable}
\end{table}

\end{document}